\documentclass[aps,prd,groupedaddress]{revtex4}
\usepackage{graphicx}
\usepackage{epsfig}
\usepackage{amsfonts}
\usepackage{amssymb}
\usepackage{epsf}
\usepackage{color}
\newcommand{\insertplot}[5]{\begin{figure}
 \hfill\hbox to 0.05in{\vbox to #5in{\vfill
 \inputplot{#1}{#4}{#5}}\hfill}
 \hfill\vspace{-.1in}
 \caption{#2}\label{#3}
 \end{figure}}
\newcommand{\inputplot}[3]{
 \special{ps: plotfile #1}
}

\newcounter{fig}

\begin{document}

\title{Scalarized nutty wormholes}
\author{Rustam Ibadov}
\email[]{ibrustam@mail.ru}
\affiliation{Department of Theoretical Physics and Computer Science, Samarkand State University,
Samarkand 140104, Uzbekistan}
\author{Burkhard Kleihaus}
\email[]{b.kleihaus@uni-oldenburg.de}
\affiliation{Institute of Physics, University of Oldenburg, D-26111 Oldenburg, Germany}
\author{Jutta Kunz}
\email[]{jutta.kunz@uni-oldenburg.de}
\affiliation{Institute of Physics, University of Oldenburg, D-26111 Oldenburg, Germany}
\author{Sardor Murodov}
\email[]{mursardor@mail.ru}
\affiliation{Department of Theoretical Physics and Computer Science, Samarkand State University,
Samarkand 140104, Uzbekistan$ $}

\date{\today}
\begin{abstract}
We construct scalarized wormholes with a NUT charge in higher curvature theories.
We consider both Einstein-scalar-Gauss-Bonnet and 
Einstein-scalar-Chern-Simons theories, following a recent paper by
Brihaye et al.~\cite{Brihaye:2018bgc}, 
where spontaneously scalarised Schwarzschild-NUT solutions were studied.
By varying the coupling parameter and the scalar charge we determine the
domain of existence of the scalarized nutty wormholes, 
and their dependence on the NUT charge.
In the Gauss-Bonnet case the known set of scalarized wormholes \cite{Antoniou:2019awm}
is reached in the limit of vanishing NUT charge.
In the Chern-Simons case, however, the limit is peculiar,
since with vanishing NUT charge the coupling constant diverges.
We focus on scalarized nutty wormholes with a single throat 
and study their properties.
All these scalarized nutty wormholes feature a critical polar angle, 
beyond which closed timelike curves are present.
\end{abstract}

\maketitle

\section{Introduction}

The fascinating phenomenon of scalarization has led to a large variety of interesting observations
in the context of compact objects.
Scalarization arises, when the generalized Einstein-Klein-Gordon equations
lead to solutions with a non-trivial scalar field, caused by the presence of
an adequate source term.
Depending on the properties of this source term distinct types of scalarized solutions arise.
When the source term in the scalar field equation does not vanish
for vanishing scalar field, all solutions will be scalarized,
and the solutions of ordinary General Relativity (GR) will not be solutions
of the coupled set of field equations.
In contrast, when the source term in the scalar field equation does vanish for vanishing
scalar field the GR solutions do remain solutions of the
generalized set of field equations. However, they develop tachyonic instabilities,
where new scalarized solutions arise.

The latter phenomenon was first observed for neutron stars in scalar-tensor theories \cite{Damour:1993hw}
where it is referred to as matter-induced spontaneous scalarization,
since the source term for the scalar field is provided by the highly compact nuclear matter.
Only much more recently it was observed that in the case of the vacuum black holes of
GR spontaneous scalarization is possible as well, when another type of source term
for the scalar field is provided. Coupling the higher curvature Gauss-Bonnet (GB) invariant
to the scalar field with an appropriate coupling function
curvature-induced spontaneously scalarized black holes arise,
representing scalarized Schwarzschild and Kerr black holes
\cite{Antoniou:2017acq,Doneva:2017bvd,Silva:2017uqg,Antoniou:2017hxj,Blazquez-Salcedo:2018jnn,Doneva:2018rou,Minamitsuji:2018xde,Silva:2018qhn,Brihaye:2018grv,
Myung:2018jvi,Bakopoulos:2018nui,Doneva:2019vuh,Myung:2019wvb,Cunha:2019dwb,
Macedo:2019sem,Hod:2019pmb,Bakopoulos:2019tvc, Collodel:2019kkx,Bakopoulos:2020dfg,
Blazquez-Salcedo:2020rhf,Blazquez-Salcedo:2020caw,Herdeiro:2020wei,Berti:2020kgk}.
Moreover, the curvature-induced spontaneous scalarized Kerr black holes come in two types,
those that arise in the slow rotation limit from the scalarized Schwarzschild black holes,
and those that do not possess a slow rotation limit \cite{Dima:2020yac,Hod:2020jjy,Doneva:2020nbb}.

An alternative higher curvature invariant to study curvature-induced spontaneously scalarized
black holes is the Chern-Simons (CS) invariant. However, in the static case of the
Schwarzschild metric the invariant vanishes, and therefore no spontaneously scalarized Schwarzschild
black holes arise. This is different for the Kerr metric, since rotation leads to a finite
CS source term for the scalar field
\cite{Yunes:2009hc,Konno:2009kg,Cambiaso:2010un,Yagi:2012ya,Stein:2014xba,Konno:2014qua,McNees:2015srl,Delsate:2018ome}, which should allow for spontaneously scalarized Kerr black holes.
In order to learn about scalarized rotating CS black holes 
without having to deal with the full complexity of the challenging set of the resulting coupled
partial differential equations,
one may first resort to the technically much simpler case and include a NUT charge
\cite{Taub:1950ez,Newman:1963,Misner:1965zz,Misner:1967,Bonnor:1969,Bonnor:2001,Manko:2005nm}
instead of rotation,
as pursued successfully by Brihaye et al.~\cite{Brihaye:2018bgc,Brihaye:2016lsx}.
In that case, a much simpler set of ordinary differential equations (ODEs) results,
since the angular dependence of the scalarised solutions factorizes.

Inspired by Brihaye et al.~\cite{Brihaye:2018bgc,Brihaye:2016lsx}, we here follow their motivation
and apply this strategy to wormholes, constructing scalarized nutty wormholes in 
higher curvature theories employing either a GB term or a CS term
as source term for the scalar field.
Following Brihaye et al.~\cite{Brihaye:2018bgc},  we here employ a quadratic coupling function.
The spontaneously scalarized Schwarzschild-NUT solutions of \cite{Brihaye:2018bgc} then represent
one of the boundaries of the domain of existence of the scalarized nutty wormholes.
However, the wormhole solutions are not spontaneously scalarized, since when 
the scalar field vanishes, pure vacuum GR is retained, and there are no Lorentzian
traversable wormhole solutions in vacuum GR (see e.g., \cite{Morris:1988cz,Visser:1995cc,Lobo:2017eum}).

To obtain traversable wormhole solutions the energy conditions must be violated.
In GR this can be achieved by the presence of exotic matter.
However, by allowing for alternative theories of gravity traversable wormholes can be obtained without
the need for exotic matter (see e.g.,
\cite{Hochberg:1990is,Fukutaka:1989zb,Ghoroku:1992tz,Furey:2004rq,Eiroa:2008hv,Bronnikov:2009az,Kanti:2011jz,Kanti:2011yv,Lobo:2009ip,Harko:2013yb}).
Employing the string theory motivated dilaton-GB coupling, static scalarized wormholes
were shown to exist, and their domain of existence and their properties were studied before
\cite{Kanti:2011jz,Kanti:2011yv,Antoniou:2019awm}.
In the dilatonic case, the black hole boundary of the domain of existence corresponds static dilatonic black holes
\cite{Kanti:1995vq}.
For other coupling functions, which give rise to spontaneously scalarized GB black holes,
the black hole boundary of the domain of existence of scalarized wormholes \cite{Antoniou:2019awm,Ibadov:2020btp}
consists of the corresponding spontaneously scalarized black holes 
\cite{Antoniou:2017acq,Doneva:2017bvd,Silva:2017uqg}.

The domain of existence of scalarized wormholes is further bounded by 
a set of solutions, where singularities are encountered, 
and by a set of solutions, where the wormhole throat becomes degenerate
\cite{Kanti:2011jz,Kanti:2011yv,Antoniou:2019awm,Ibadov:2020btp}.
In the latter case, this degeneracy reveals, that in addition to
wormholes with a single throat also wormholes with an equator
and a double throat exist. The throat(s) and equator arise naturally
in these solutions in one of the two parts of the spacetime.
However, when simply continuing the solutions beyond the throat (or equator)
a singularity will invariably be encountered.
In order to obtain wormholes without such singularities, 
symmetry has been imposed  with respect to the throat (or equator).
This entails that a thin shell of matter is needed
at the throat (or equator), to satisfy the respective
Israel junction conditions \cite{Israel:1966rt,Davis:2002gn}.
Typically, this matter can be ordinary matter,
thus no exotic matter is needed
\cite{Kanti:2011jz,Kanti:2011yv,Antoniou:2019awm,Ibadov:2020btp}.

Here we generalize these scalarized wormhole solutions in two ways.
On the one hand, we include a NUT charge 
and on the other hand we consider besides the GB invariant also the CS invariant.
The presence of the NUT charge implicates a Misner string on the polar axis.
Therefore the resulting spacetimes are not asymptotically flat in the usual sense.
However, all unknown functions of the wormhole solutions are only functions of the radial coordinate,
and their asymptotic fall-off is of the usual type of an asymptotically flat spacetime,
as in the case of scalarized Schwarzschild-NUT solutions \cite{Brihaye:2018bgc,Brihaye:2016lsx}.
As always, the presence of a NUT charge gives rise to closed timelike curves,
since the metric component of the azimuthal angle, $g_{\varphi\varphi}$,
changes sign in the manifold at a critical value of the polar angle $\theta$.
Here we show that all scalarized nutty wormhole solutions possess such a critical polar angle
at their throat. In the black hole limit the throat changes into a horizon,
where the critical polar angle goes to zero.

The paper is organized as follows: In section II we present the actions involving
the GB and the CS term and exhibit the equations of motion for both cases.
We then discuss the boundary conditions, the conditions for the center,
i.e., the throat (or equator),
the junction conditions, and the null energy condition (NEC).
Subsequently we address the numerical procedure and present our results in section III.
These include, in particular, the profile functions for the scalarized nutty wormhole solutions,
the violation of the NEC,
the domain of existence with its outer boundaries, and 
an analysis of the junction conditions for the thin shell of matter at the throat.
We give our conclusions and an outlook in section IV.

\section{Theoretical setting}

\subsection{Action and equations of motion}

We consider the effective action for Einstein-scalar-higher curvature invariant theories
\begin{eqnarray}  
S=\frac{1}{16 \pi}\int \left[R - \frac{1}{2}
 \partial_\mu \phi \,\partial^\mu \phi 
 + F(\phi){\cal I}(g) \right] \sqrt{-g} d^4x  \ ,
\label{act}
\end{eqnarray} 
where $R$ is the curvature scalar,
and $\phi$ denotes the scalar field, 
that is coupled with some coupling function $F(\phi)$ to an invariant
${\cal I}(g)$.  
For the coupling function $F(\phi)$ we choose a quadratic
$\phi$-dependence with coupling constant $\alpha$,
\begin{equation}
F(\phi) = \alpha \phi^2 .
\label{FandU}
\end{equation}
For the invariant ${\cal I}(g)$ we make two choices, (i) the Gauss-Bonnet term
\begin{eqnarray} 
{\cal I}(g)= R^2_{\rm GB} = R_{\mu\nu\rho\sigma} R^{\mu\nu\rho\sigma}
- 4 R_{\mu\nu} R^{\mu\nu} + R^2 \ , 
\end{eqnarray} 
and (ii) the Chern-Simons term
\begin{eqnarray} 
{\cal I}(g)= R^2_{\rm CS} = 
   ^\ast\!{\!\!R^\mu_{{\phantom \mu}\nu}}^{\rho\sigma} R^\nu_{{\phantom \nu}\mu\rho\sigma}
\ , 
\end{eqnarray} 
where the Hodge dual of the Riemann-tensor 
$^\ast{\!R^\mu_{{\phantom \mu}\nu}}^{\rho\sigma} 
  = \frac{1}{2}\eta^{\rho\sigma\kappa\lambda} R^\mu_{{\phantom \mu}\nu\kappa\lambda}
$
is defined with the 4-dimensional Levi-Civita tensor 
$\eta^{\rho\sigma\kappa\lambda}$.
While both invariants are topological in four dimensions, 
the coupling to the scalar field $\phi$ 
via the coupling function $F(\phi)$ provides
significant contributions to the equations of motion.

We obtain the coupled set of field equations
by varying the action (\ref{act}) with respect to 
the scalar field and to the metric, 
\begin{equation}
\nabla^\mu \nabla_\mu \phi  + \frac{dF(\phi)}{d\phi}{\cal I}=0 \ ,
\label{scleq}
\end{equation}
\begin{equation}
G_{\mu\nu}  = \frac{1}{2}T^{({\rm eff})}_{\mu\nu} \ , 
\label{Einsteq}
\end{equation}
where $G_{\mu\nu}$ is the Einstein tensor and $T^{({\rm eff})}_{\mu\nu}$
denotes the effective stress-energy tensor	
\begin{equation}
T^{({\rm eff})}_{\mu\nu} = T^{(\phi)}_{\mu\nu} +  T^{({\cal I})}_{\mu\nu} \ ,
\label{teff}
\end{equation}
which consists of the scalar field contribution
\begin{equation}
T^{(\phi)}_{\mu\nu} = \left(\nabla_\mu \phi\right) \left(\nabla_\nu \phi\right)
                   -\frac{1}{2}g_{\mu\nu}\left(\nabla_{\!\rho}\, \phi\right) \left(\nabla^\rho \phi\right) \ , 
\label{tphi}
\end{equation}
and a contribution from the respective invariant ${\cal I}(g)$. 
For the chosen invariants we obtain (i)
\begin{equation}
T^{({\cal I})}_{\mu\nu} =
\frac{1}{2}\left(g_{\rho\mu} g_{\lambda\nu}+g_{\lambda\mu} g_{\rho\nu}\right)
\eta^{\kappa\lambda\alpha\beta}\tilde{R}^{\rho\gamma}_{\phantom{\rho\gamma}\alpha\beta}
\nabla_\gamma \nabla_\kappa F(\phi) \ ,
\label{teffi}
\end{equation}
where
$\tilde{R}^{\rho\gamma}_{\phantom{\rho\gamma}\alpha\beta}=\eta^{\rho\gamma\sigma\tau}
R_{\sigma\tau\alpha\beta}$ and $\eta^{\rho\gamma\sigma\tau}= 
\epsilon^{\rho\gamma\sigma\tau}/\sqrt{-g}$,
and (ii) 
\begin{equation}
T^{({\cal I}) \mu\nu} =
-8 \left[\nabla_\rho F(\phi)\right] \epsilon^{\rho\sigma\tau ( \mu}
                  \nabla_\tau R^{\nu )}_{\phantom{\nu )}\sigma}
                 +\left[\nabla_\rho \nabla_\sigma F(\phi)\right]
^\ast{\!\!R^{\sigma ( \mu \nu )\rho }}	  .
\label{teffii}
\end{equation}

To obtain static, spherically symmetric wormhole solutions with a NUT charge $N$ we assume 
the line element to be of the form 
\begin{equation}
ds^2 = -e^{f_0}\left(dt - 2 N \cos\theta d\varphi\right)^2 +e^{f_1}\left[dr^2 
+r^2\left( d\theta^2+\sin^2\theta d\varphi^2\right) \right]\ ,
\label{met}
\end{equation}
All three functions,
the two metric functions $f_0$ and $f_1$ and the scalar field function $\phi$,
depend only on the radial coordinate $r$. 

When we insert the above ansatz (\ref{met}) for the metric and the scalar field into
the scalar-field equation (\ref{scleq})
and the Einstein equations (\ref{Einsteq}) with effective 
stress-energy tensor (\ref{teff}),
we obtain five coupled, nonlinear ODEs.
However, these are not independent, since the $\theta$-dependence factorizes,
and one ODE can be treated as a constraint.
This leaves us with three coupled ODEs of second order.
Note that in case (i) 
the system can be reduced to one first order and two second order ODEs.

Inspection of the field equations reveals an invariance under the scaling transformation
\begin{equation}
r \to \chi r \ , \ \ \  N \to \chi N \ , \ \ \   t \to \chi t \ , \ \ \  F \to \chi^2 F\  , \ \ \  \chi > 0 \ .
\label{scalinvar}
\end{equation}

\subsection{Throats, equators, and boundary conditions}

In order to obtain scalarized nutty wormhole solutions, we need to 
impose an appropriate set of boundary conditions for the ODEs,
which we now address.
%
We first introduce the circumferential (or spherical) radius
\begin{equation}
R_C=e^{\frac{f_1}{2}} r \ 
\label{rc}
\end{equation}
of the wormhole spacetimes, which may
possess one or more finite extrema.
If there is a single finite extremum, 
this corresponds to the single throat of the respective wormholes.
Here we will mainly consider such single throat wormholes,
thus featuring a single minimum.
But wormholes with more extrema may also exist. 
They might, for instance, possess a local maximum surrounded by two minima.
The local maximum would then correspond to their equator, 
while the two minima would represent their two throats,
making them double throat wormholes.

To obtain the first set of boundary conditions we therefore
require the presence of an extremum 
of the spherical radius at some $r=r_0$.
This yields 
\begin{equation}
\left. \frac{dR_C}{dr} \right|_{r=r_0} = 0 \ \ \ 
\Longleftrightarrow \ \ \ \left.   \frac{df_1}{dr}\right|_{r=r_0} =-\frac{2}{r_0} \ .
\label{extr_rc}
\end{equation}
In the following we will refer to the two-dimensional submanifolds
defined by $r=r_0$
and $t=$const.~as the center of the configurations.

We note that the presence of a NUT charge leads
to an interesting feature of these wormholes. 
Unlike the usual case, the metric of these nutty wormholes 
is not Riemannian on the throat,
\begin{equation}
ds^2_{\rm th} = e^{f_1(r_0)} r_0^2 \left(d\theta^2 
                +\left[\sin^2\theta -\frac{4N^2}{r_0^2} e^{f_0(r_0)-f_1(r_0)} \cos^2\theta\right]d\varphi^2
		\right) \ .
\label{met_th}
\end{equation}
This is a consequence of the non-causal structure of 
a spacetime in the presence of a NUT charge $N$.
In this connection we introduce a critical angle $\theta_c$,
which we obtain from the condition 
$\det(g_{\rm th}) \geq 0$,
which requires $\theta_c \leq \theta \leq \pi-\theta_c$
with 
\begin{equation}
\theta_c= \left.
\arctan\left(\frac{2|N|}{r_0} e^{\frac{f_0-f_1}{2}}\right)
\right|_{r_0} .
\label{thetac}
\end{equation}

We obtain the second set of boundary conditions 
by requiring the usual boundary conditions for $r \to \infty$ \cite{Brihaye:2018bgc}.
The associated asymptotic expansions of the metric functions and the
scalar field read
%
%
\begin{eqnarray}
f_0  & = & - \frac{2M}{r} + {\cal O} \left(r^{-3}\right) ,
\label{expf0} \\
f_1  & = &  \ \frac{2M}{r} + {\cal O} \left(r^{-2}\right) ,
\label{expf1} \\
\phi  & = & \phi_\infty - \frac{D}{r} + {\cal O} \left(r^{-3}\right) ,
\label{exphi0} 
\end{eqnarray}
where $M$ denotes the mass of the wormholes
and $D$ corresponds to their scalar charge. 
The quantity $\phi_\infty$ represents the asymptotic
value of the scalar field, 
which we set to zero in our analysis, $ \phi_\infty =0$.

\subsection{Junction conditions}
In order to obtain wormholes whose geometry is symmetric with respect to the center
and which do not possess any singularity (apart from the Misner string),
we need to impose junction conditions at the center.
For the discussion of the junction conditions it is useful to introduce
the radial coordinate $\eta$,
\begin{equation}
\eta= r_0\left(\frac{r}{r_0}-\frac{r_0}{r}\right) \ ,
\label{defeta}
\end{equation} 
where $r_0$ is a constant. We then define the constant $\eta_0$ via
$\eta_0 = 2 r_0$.
In terms of the new radial coordinate $\eta$ the metric reads
\begin{equation}
ds^2 = -e^{f_0}\left(dt-2N \cos\theta d\varphi\right)^2
+e^{{F}_1}\left[d\eta^2 
               +\left(\eta^2+\eta_0^2\right) \left(d\theta^2+\sin^2\theta d\varphi^2\right)\right] \ ,
\label{ds2eta}
\end{equation} 
where we have introduced the new metric function $F_1$,
$$
e^{{F}_1} = e^{f_1}\left(1+\frac{r_0^2}{r^2}\right)^{-2} \ .
$$
Thus the center is located at $\eta=0$, and the solution on the $\eta \le 0$ part of the
manifold can be obtained from the solution on 
the $\eta \ge 0$ part of the manifold by imposing the symmetry conditions
$f_0(-\eta) = f_0(\eta)$, ${F}_1(-\eta) = {F}_1(\eta)$ and
$\phi(-\eta) = \phi(\eta)$.
However, these conditions generically introduce jumps in the derivatives of the functions $f_0$ and $\phi$
at the center $\eta=0$, which may be attributed to a thin shell of matter that is localized at the center.

To properly embed such a thin shell of matter in the complete wormhole solution, we make use of
an appropriate set of junction conditions \cite{Israel:1966rt,Davis:2002gn}.
In particular, we consider jumps in the coupled set of Einstein and scalar field equations 
that arise when $\eta \rightarrow -\eta$, 
\begin{equation}
\langle G^\mu_{\phantom{a}\nu} -T^\mu_{\phantom{a}\nu}\rangle = s^\mu_{\phantom{a}\nu} \ , \ \ \ 
\langle \nabla^2 \phi + \dot{F} {\cal I}\rangle = s_{\rm scal} \ ,
\label{jumps}
\end{equation}
%
where we have denoted the stress-energy tensor of the matter at the center by $s^\mu_{\phantom{a}\nu}$,
and the source term for the scalar field by $s_{\rm scal}$.
We would like the matter forming the thin shells
to be some form of ordinary (non-exotic) matter.
We will therefore assume that there is a perfect fluid at the center
which has pressure $p$ and energy density $\epsilon_c$, and that there is 
a scalar charge density $\rho_{\rm scal}$ together with a gravitational source
\cite{Kanti:2011jz,Kanti:2011yv}
\begin{equation}
S_\Sigma = \int \left[\lambda_1 + 2 \lambda_0 F(\phi) \bar{R}\right]\sqrt{-\bar{h}} d^3 x \ .
\label{act_th}
\end{equation}
Here we have introduced the constants $\lambda_1$ and $\lambda_0$, $\bar{h}_{ab}$ denotes the
induced metric at the center, and $\bar{R}$ denotes the associated Ricci scalar. 
In order to obtain the junction conditions, we substitute the metric into the sets of equations,
and we introduce the abbreviation $dF(\phi)/d\phi = \dot F$ for the derivative of the coupling function.

Now we derive the junction conditions for both invariants separately.
Note, that here and in the following all functions and derivatives are evaluated
at the center.
In the case of the Gauss-Bonnet invariant we find the equations
\begin{eqnarray}
\frac{4}{\eta_0^2} \dot{F} \phi' 
\left(\eta_0^2 e^{-\frac{3}{2} {F}_1} +3 N^2 e^{f_0-\frac{5}{2} {F}_1} \right) 
& = &
\lambda_1\eta_0^2 + 4\lambda_0 F \frac{ \eta_0^2 e^{-{F}_1}
                               +3 N^2 e^{f_0-2{F}_1}}{\eta_0^2}
- \epsilon_c\eta_0^2 \ ,  
\label{junc_gb_00}\\
N\cos\theta\left[  
\eta_0^2 f_0' e^{-\frac{{F}_1}{2}}
-8\dot{F} \phi' 
\left(e^{-\frac{3}{2} {F}_1}  +\frac{4 N^2}{\eta_0^2}  e^{f_0-\frac{5}{2} {F}_1}\right)
\right]
& = &
2 N \cos\theta \left[  
\left(\epsilon_c + p\right) \eta_0^2 
- 4\lambda_0 F \frac{ \eta_0^2 e^{-{F}_1}
                               +4 N^2 e^{f_0-2{F}_1}}{\eta_0^2}
\right] \ ,
\label{junc_gb_0p}\\
\frac{\eta_0^2 f_0'}{2} e^{-\frac{{F}_1}{2}}
                       -\frac{4 N^2}{\eta_0^2} \dot{F} \phi' e^{f_0-\frac{5}{2} {F}_1}
& = &
p \eta_0^2 +\lambda_1  \eta_0^2
- 4\lambda_0 N^2  F\frac{e^{f_0-2{F}_1}}{\eta_0^2} \ ,
\label{junc_gb_pp}\\
e^{-{F}_1}\phi' - 4 \frac{\dot{F}}{\eta_0^4} f_0' 
\left(\eta_0^2 e^{-2 {F}_1}+3 N^2 e^{ f_0-3{F}_1} \right)
& = &
-4\lambda_0\frac{\dot{F}}{\eta_0^4}\left(\eta_0^2 e^{-{F}_1} 
                                      + N^2 e^{f_0-2{F}_1}\right)
+\frac{\rho_{\rm scal}}{2} \ ,
\label{junc_gb_ph}
\end{eqnarray}
which follow from the 
$\left( ^t_{\phantom{a}t}\right)$, 
$\left( ^t_{\phantom{a}\varphi}\right)$, and
$\left( ^\varphi_{\phantom{a}\varphi}\right)$
components of the Einstein equations and from the scalar field equation, respectively.
Note that the $\left( ^\theta_{\phantom{a}\theta}\right)$ equation is equivalent to the 
$\left( ^\varphi_{\phantom{a}\varphi}\right)$ equation, and that all other equations are
satisfied trivially.
We also remark that the $\theta$ dependence in the 
$\left( ^t_{\phantom{a}\varphi}\right)$ equation
factorizes, and that this equation is satisfied once 
the $\left( ^t_{\phantom{a}t}\right)$ and $\left( ^\varphi_{\phantom{a}\varphi}\right)$
equations are solved.

As an example we consider pressureless matter, $p=0$.
With 
\begin{equation}
\lambda_0 = \frac{\dot{F}}{F} e^{-\frac{{F}_1}{2}} \phi'\ , \ \ \ 
\lambda_1 =\frac{ f_0'}{2}e^{-\frac{{F}_1}{2}}\ ,
\label{lambda_ex_gb}
\end{equation}
we find the simple result
\begin{equation}
\epsilon_c = 
\frac{f_0'}{2}e^{-\frac{{F}_1}{2}} \ .	     
\label{eden_ex_gb}
\end{equation}
Since for our solutions $f_0' > 0$ the energy density $\epsilon_c$ is always positive for this
choice of the constants $\lambda_0$ and $\lambda_1$.

We now turn to case of the Chern-Simons invariant where we
find the equations
\begin{eqnarray}
8N  \dot{F} \phi' f_0' e^{\frac{f_0}{2}}  e^{-2 {F}_1}
& = &
\lambda_1\eta_0^2 + 4\lambda_0 F \frac{ \eta_0^2 e^{-{F}_1}
                               +3 N^2 e^{f_0-2{F}_1}}{\eta_0^2}
- \epsilon_c\eta_0^2 \ ,  
\label{junc_cs_00}\\
N \cos\theta f_0' \left(\eta_0^2 e^{-\frac{{F}_1}{2}}
                       -24 N \dot{F} \phi' e^{\frac{f_0}{2}-2{F}_1}\right)
& = &
2 N \cos\theta \left(\epsilon_c + p\right) \eta_0^2 
- 8 N \cos\theta \lambda_0 F \frac{ \eta_0^2 e^{-{F}_1}
                               +4 N^2 e^{f_0-2{F}_1}}{\eta_0^2} \ ,
\label{junc_cs_0p}\\
\frac{f_0'}{2} \left(\eta_0^2 e^{-\frac{{F}_1}{2}}
                       -8 N \dot{F} \phi' e^{\frac{f_0}{2}-2{F}_1}\right)
& = &
p \eta_0^2 +\lambda_1  \eta_0^2
- 4\lambda_0 N^2 F \frac{e^{f_0-2{F}_1}}{\eta_0^2} \ ,
\label{junc_cs_pp}\\
e^{-{F}_1}\phi' - 4 N \frac{\dot{F}}{\eta_0^2} (f_0')^2 
e^{\frac{f_0-5 {F}_1}{2}}
& = &
-4\lambda_0\frac{\dot{F}}{\eta_0^4}\left(\eta_0^2 e^{-{F}_1} 
                                      +N^2 e^{f_0-2{F}_1}\right)
+\frac{\rho_{\rm scal}}{2} \ ,
\label{junc_cs_ph}
\end{eqnarray}
which follow again from the 
$\left( ^t_{\phantom{a}t}\right)$, 
$\left( ^t_{\phantom{a}\varphi}\right)$, 
$\left( ^\varphi_{\phantom{a}\varphi}\right)$
components of the Einstein equations and the scalar field equation, respectively.

Again we consider as an example pressureless matter, $p=0$.
With 
\begin{equation}
\lambda_0 = \frac{\eta_0^2 \dot{F}}{N F} e^{-\frac{f_0}{2}} \ , \ \ \ 
\lambda_1 =-\frac{ f_0'}{2}e^{-\frac{{F}_1}{2}}\ ,
\label{lambda_ex_cs}
\end{equation}
we now find 
\begin{equation}
\epsilon_c = 
\frac{f_0'}{2 N}\left(N e^{-\frac{{F}_1}{2}} 
                     + 8\dot{F}\phi'e^{-\frac{f_0}{2}-{F}_1}\right)
       +\frac{4 N}{\eta_0^2}\dot{F}\phi'f_0' e^{\frac{f_0}{2}-2{F}_1}		  \ .   
\label{eden_ex_cs}
\end{equation}

\subsection{Energy conditions}

In wormhole solutions the null energy condition (NEC)
\begin{equation}
T_{\mu\nu} n^\mu n^\nu \geq 0 \ 
\label{NEC}
\end{equation}
must be violated, where
$n^\mu$ is any null vector ($n^\mu n_\mu=0$). 
Thus it is sufficient to show that null vectors exist, such that 
$T_{\mu\nu} n^\mu n^\nu < 0$ in some spacetime region.
Such a null vector $n^\mu$ is given by
\begin{equation}
n^\mu=\left(1,\sqrt{-g_{tt}/g_{\eta\eta}},0,0\right) \ ,
\end{equation}
and thus $n_\mu=\left(g_{tt},\sqrt{-g_{tt}\,g_{\eta\eta}},0,0\right)$.
The NEC then takes the form
\begin{equation}
T_{\mu\nu}n^\mu n^\nu=T^t_t n^t n_t + T^\eta_\eta n^\eta n_\eta
=-g_{tt}\,(-T^t_t +T^\eta_\eta)   \ .
\end{equation}
Consequently the NEC is violated when 
\begin{equation}
 -T_t^t + T_\eta^\eta  <  0 \ .
\label{nec1}
\end{equation}
Alternatively, considering the null vector
\begin{equation}
n^\mu=\left(1,0,\sqrt{-g_{tt}/g_{\theta \theta}},0\right) \ ,
\end{equation}
the NEC is violated when
\begin{equation}
-T_t^t + T_\theta^\theta < 0 \ . 
\label{nec2}
\end{equation}
These conditions have been addressed before for various scalarized wormhole solutions 
\cite{Morris:1988cz,Kanti:2011jz,Kanti:2011yv,Antoniou:2019awm}.

\section{Results}

\subsection{Numerics}

In order to solve the coupled Einstein and scalar field equations numerically we 
introduce the inverse radial coordinate $x=1/r$.
The asymptotic region $r\to \infty$
then corresponds to $x\to 0$. 
In this region the expansion of the metric functions and the
scalar field reads (see Eqs.~(\ref{expf0})-(\ref{exphi0}))
%
\begin{equation}
f_0  = - 2 M x + {\cal O}{x^3} , \ \ \ 
f_1   =   2 M x + {\cal O}{x^2} , \ \ \
\phi   =  \phi_\infty - D x + {\cal O}{x^3} .
\label{exphi} 
\end{equation}
We treat the system of ODEs as an initial value problem,
for which we employ the fourth order Runge Kutta method.
From the above expansion we read off the initial values,
%
%
\begin{equation}
f_{0,{\rm ini}}= 0\ , \ \ \ f'_{0,{\rm ini}}=- 2M\ , \ \ \ 
f_{1,{\rm ini}}= 0\ , \ \ \ f'_{1,{\rm ini}}= 2M\ , \ \ \ 
\phi_{{\rm ini}}= 0\ , \ \ \ \phi'_{{\rm ini}}= -D \ .
\label{inicond} 
\end{equation}
The computation then starts at spatial infinity, $x=0$, 
and ends at the center at some finite $x=x_0$, where the condition
(\ref{extr_rc}) is reached.

\subsection{Solutions}

\begin{figure}[t!]
\begin{center}
(a)\includegraphics[width=.45\textwidth, angle =0]{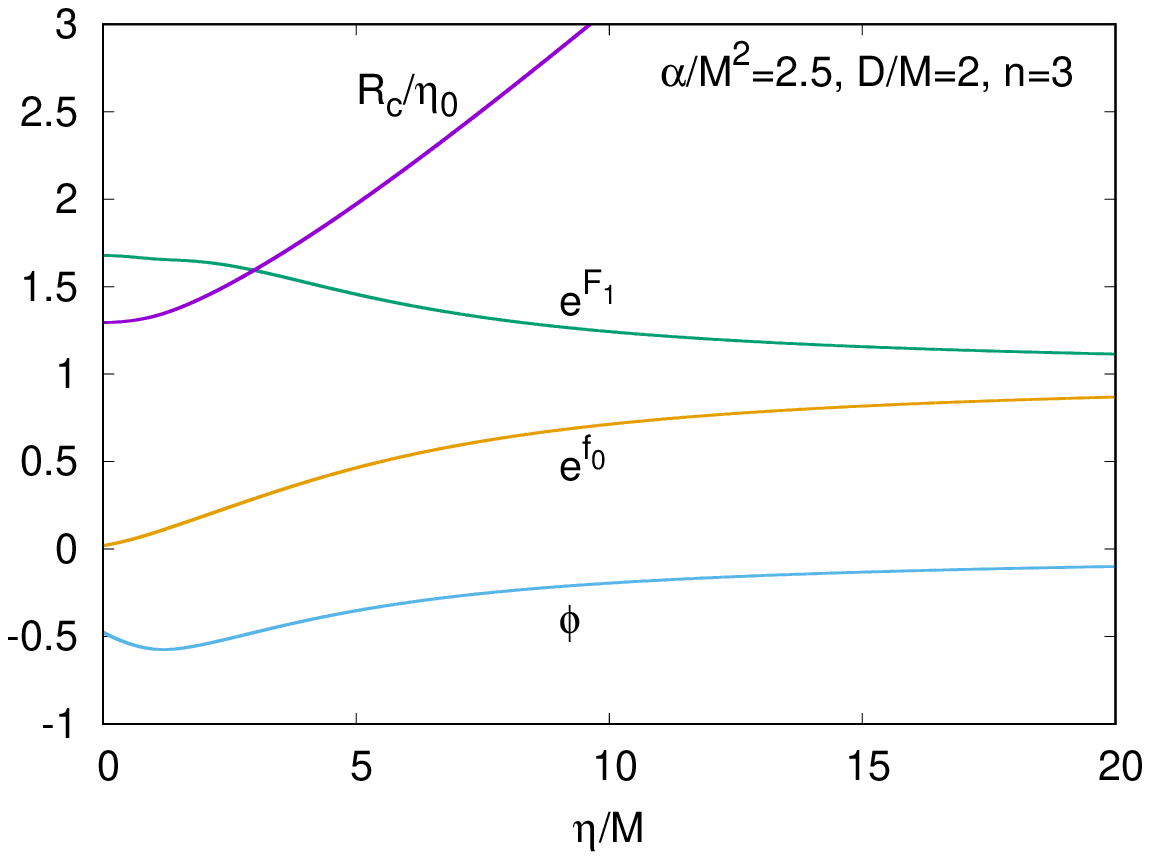}
(b)\includegraphics[width=.45\textwidth, angle =0]{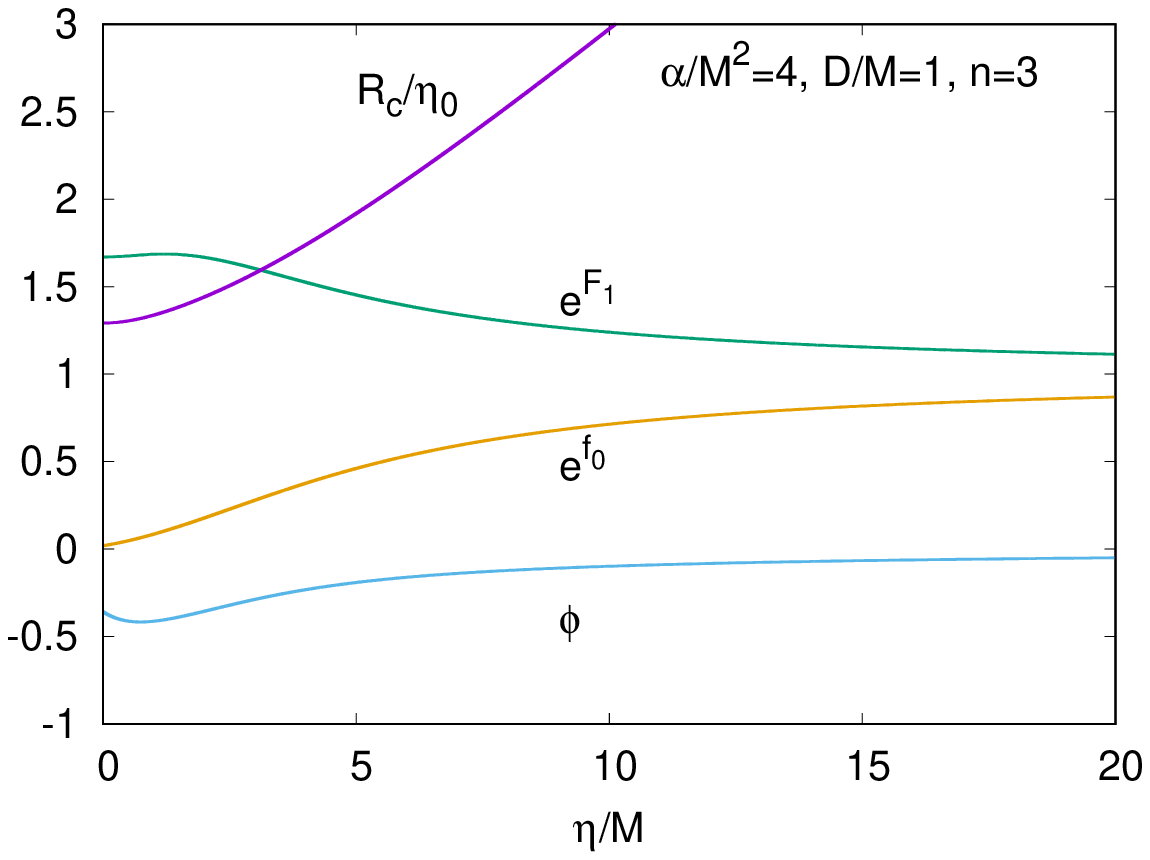}
\\
(c)\includegraphics[width=.45\textwidth, angle =0]{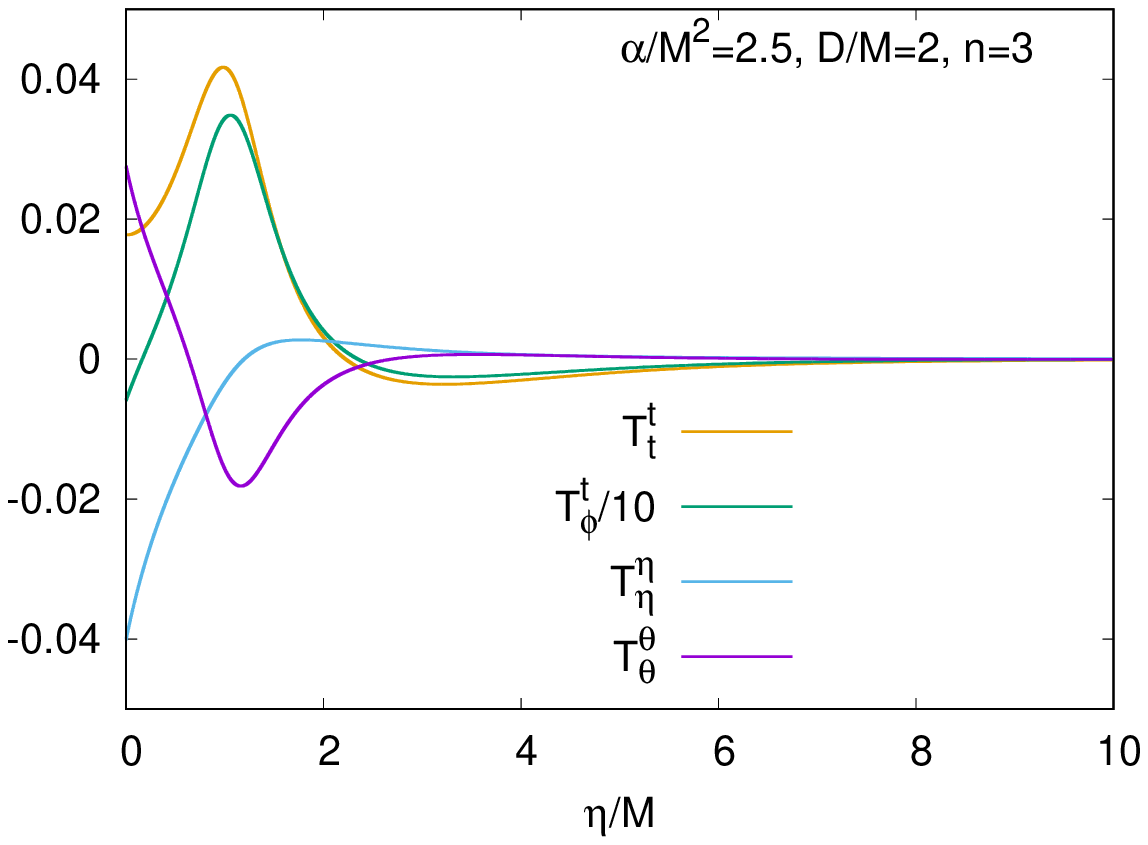}
(d)\includegraphics[width=.45\textwidth, angle =0]{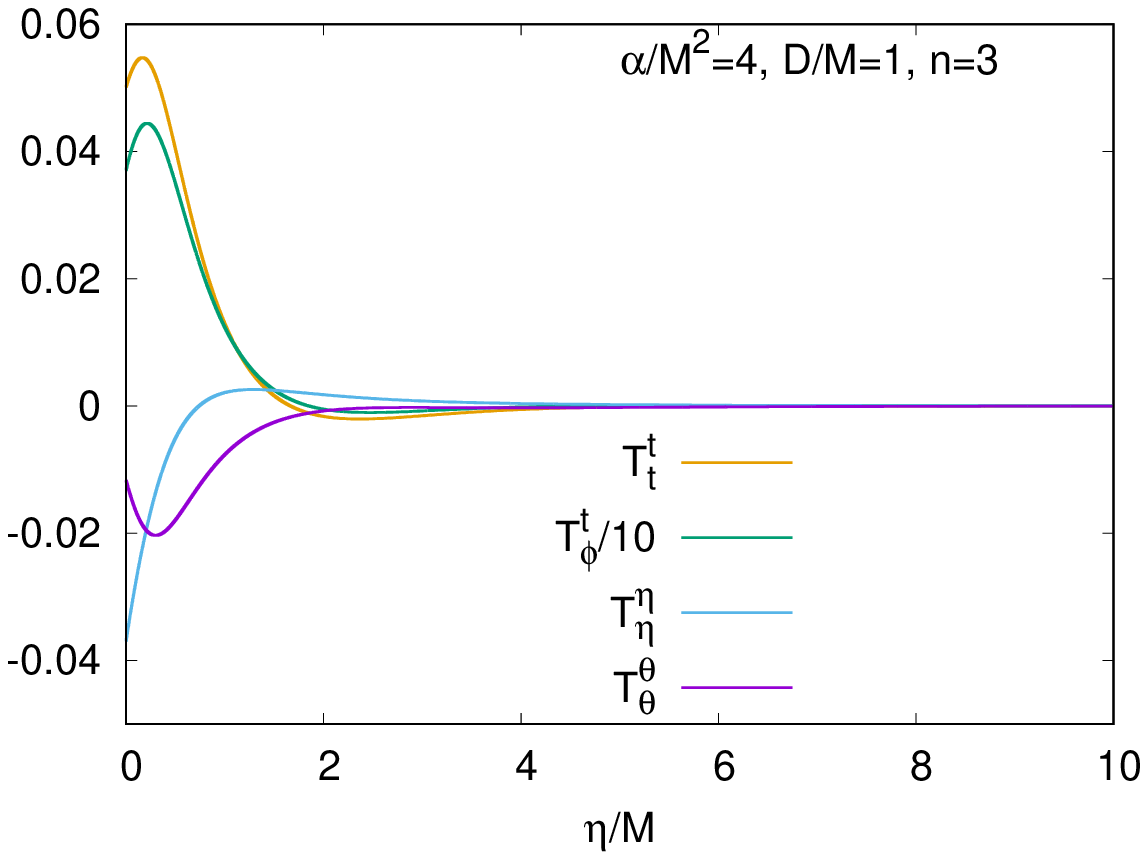}
\\
(e)\includegraphics[width=.45\textwidth, angle =0]{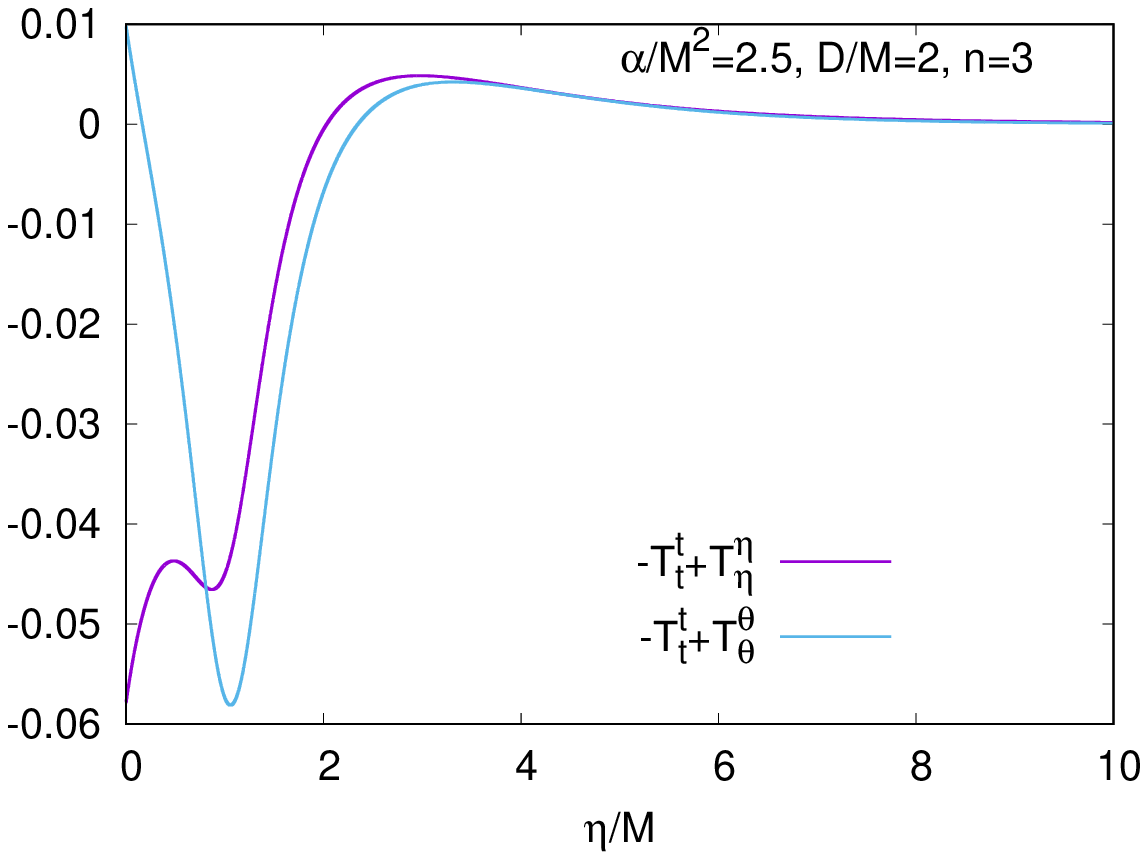}
(f)\includegraphics[width=.45\textwidth, angle =0]{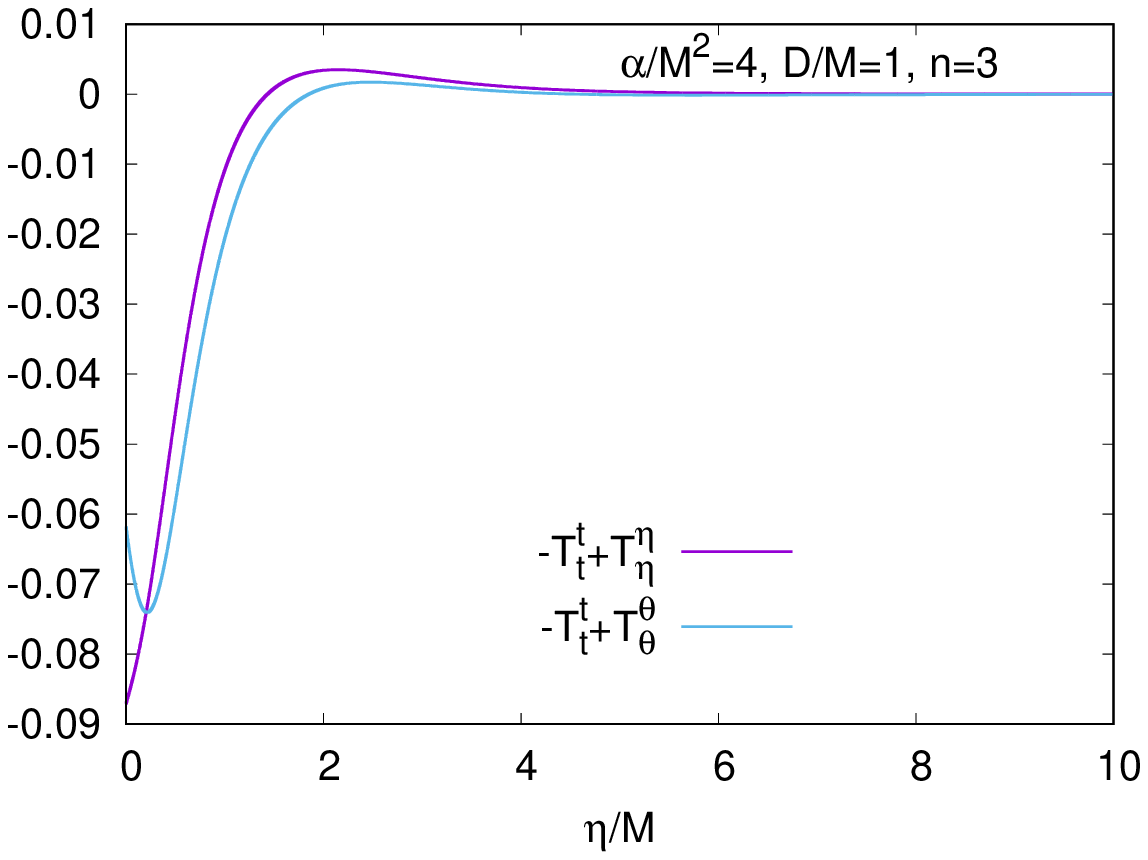}
\end{center}
\caption{
Examples of nutty wormhole solutions
(left plots: Gauss-Bonnet with parameters
$\alpha/M^2=2.5$, $D/M=2$ and $n=N/M=3$,
right plots: Chern-Simons with $\alpha/M^2=4$, $D/M=1$ and $n=N/M=3$):
(a) and (b) metric profile functions $e^{f_0}$, $e^{F_1}$, scalar field function $\phi$,
and scaled circumferential radius $R_c/\eta_0$ vs radial coordinate $\eta$;
(c) and (d) stress-energy tensor components $T^t_t$, $T^t_\phi$, $T^\eta_\eta$, and $T^\theta_\theta$
vs radial coordinate $\eta/M$;
(e) and (f) NEC conditions 
$-T_t^t + T_\eta^\eta \geq 0$ and $-T_t^t + T_\theta^\theta \geq 0$
vs radial coordinate $\eta$.
}
\label{fig_sol}
\end{figure}

By following the above numerical procedure we obtain the sets of 
nutty wormhole solutions for both invariants ${\cal I}(g)$.
Here we demonstrate some typical solutions for both cases.
We exhibit in Figs.~\ref{fig_sol} 
the metric profile functions $e^{f_0}$, $e^{F_1}$, and the
scalar field function $\phi$ versus the radial coordinate $\eta/M$
for the GB invariant (a) and the CS invariant (b),
choosing parameters $\alpha/M^2=2.5$, $D/M=2$ and $n=N/M=3$, and
$\alpha/M^2=4$, $D/M=1$ and $n=N/M=3$, respectively.
The figures also show the circumferential radius $R_c/\eta_0$
versus $\eta/M$ ($\eta_0/M=0.517$ GB invariant, $\eta_0/M=0.495$ CS invariant). As required, $R_c$ reaches an extremum 
at the center $\eta=0$.
We note that the solutions have rather similar properties
for both invariants.

To see that the wormhole solutions violate the energy conditions,
we inspect the components of the effective energy momentum tensor,
$T^t_t$, $T^t_\phi$, $T^\eta_\eta$, and $T^\theta_\theta$.
These are shown for the same solutions 
and the GB and CS invariants
in Figs.~\ref{fig_sol}(c) and (d), respectively.
In particular, we note, that the component $T^\eta_\eta$
is negative in the vicinity of the center for both invariants.
Moreover, all components are negative in some region of the spacetime.
We exhibit in
Figs.~\ref{fig_sol}(e) and (f) the NEC conditions
$ -T_t^t + T_\eta^\eta \geq 0$
and
$-T_t^t + T_\theta^\theta \geq 0$
for the GB and CS invariants, respectively.
The figures clearly demonstrate the NEC violation for both invariants.

\subsection{Domain of existence}

\begin{figure}[t!]
\begin{center}
\includegraphics[width=.45\textwidth, angle =0]{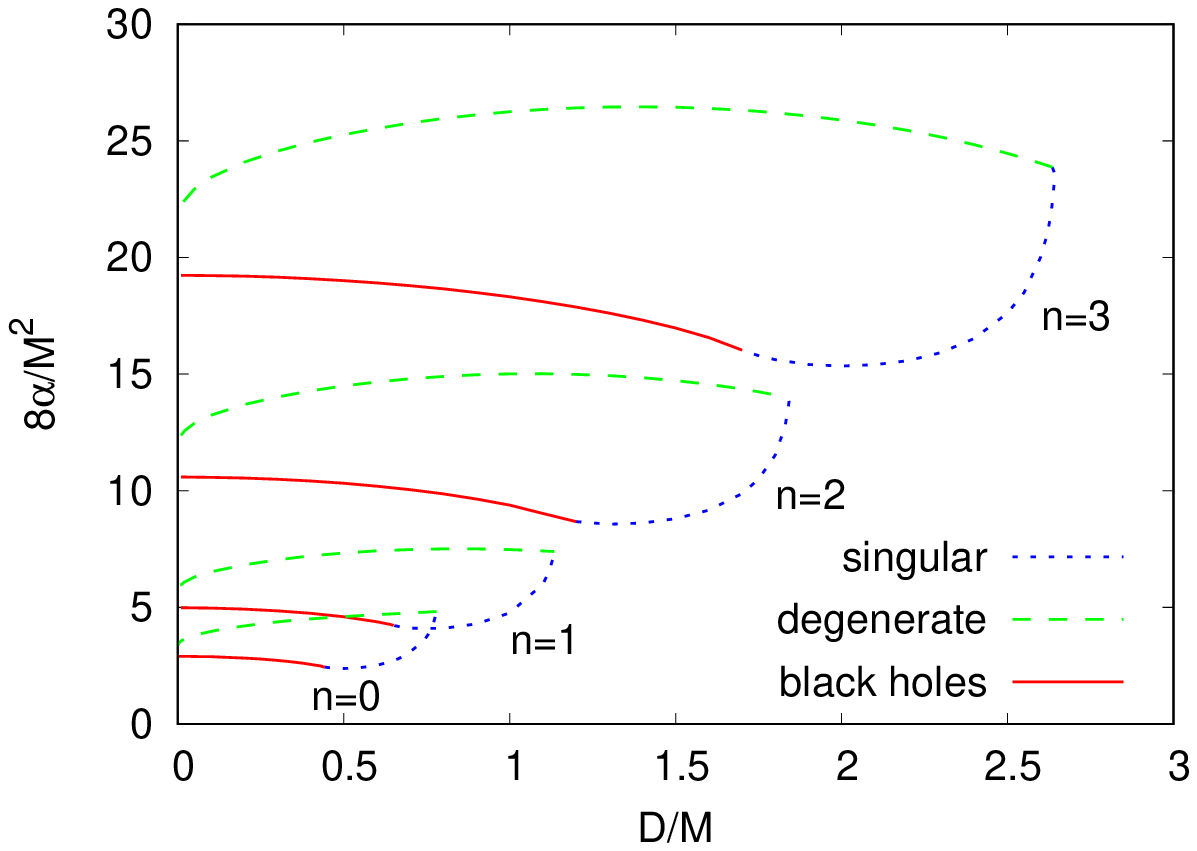}
\includegraphics[width=.45\textwidth, angle =0]{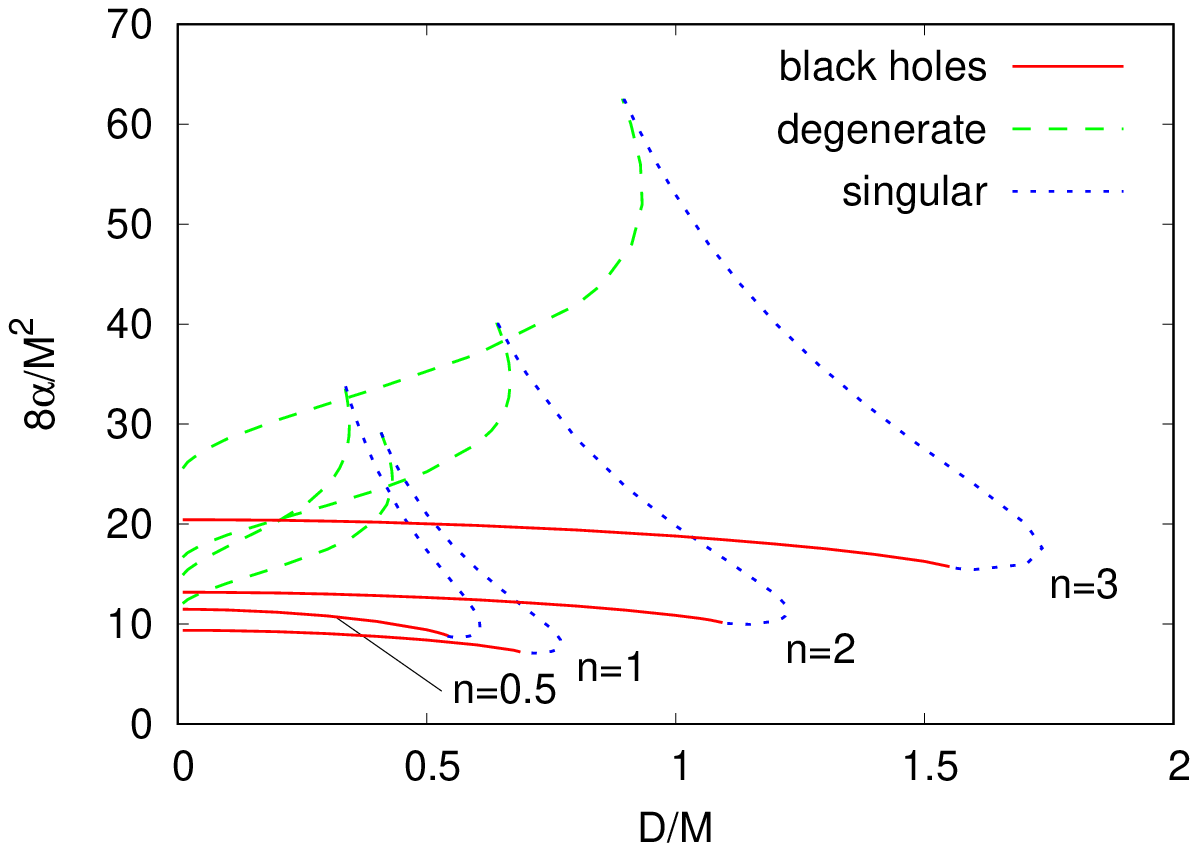}
\end{center}
\caption{
Domain of existence
(left plot: Gauss-Bonnet, right plot: Chern-Simons)
for several values of the scaled NUT charge $n=N/M$:
scaled coupling constant $\alpha/M^2$ vs
scaled scalar charge $D/M$.
The solid red curves represent the black hole limit,
the dashed green curves the degenerate wormhole limit,
and the dotted blue curves the singular limit.
}
\label{fig_dom1}
\end{figure}

We now address the domain of existence of these nutty wormhole solutions.
We exhibit the domain of existence in Fig.~\ref{fig_dom1}
for the GB (a) and CS (b) invariants, for a set of values of
the scaled NUT charge $n=N/M$.
In particular, we show the outer boundaries of the 
respective domains of existence, by presenting the scaled coupling
constant $\alpha/M^2$ versus the scaled scalar charge $D/M$.

The existence of the wormhole solutions requires the presence of a
non-trivial scalar field. Since the equations are invariant under
the transformation $\phi \to - \phi$, the domain of existence is
symmetric with respect to $D \to -D$, and it is sufficient to
only exhibit $D\ge 0$.
For $D=0$ the scalar field vanishes, and thus $D=0$ represents
the trivial boundary of the domain of existence,
where pure GR solutions reside.
The first non-trivial boundary corresponds to scalarized nutty black holes,
and is shown by the solid red curves. These scalarized black holes
were obtained before by Brihaye et al.~\cite{Brihaye:2018bgc}.
Here the center/throat turns into a black hole horizon.

The second non-trivial boundary in the figures is termed degenerate
and shown by the dashed green curves.
To understand this boundary, we recall the numerical construction
of the solutions.
The calculation is ended, when an extremum is reached.
However, in principle, we can continue the calculation
beyond the extremum, where we might find a second extremum.
The first one then corresponds to a throat
while the second one corresponds to an equator.
As the coupling constant is varied, the two extrema
will approach each other until finally a degenerate
extremum is reached.
This third boundary represents precisely the values
of the coupling constant, where such a degenerate
extremum is reached.

The last boundary has been labeled singular,
and is shown by the dotted blue curves.
At this boundary the calculations reveal the appearance
of a singularity somewhere in the spacetime,
that is of the cusp type
(see \cite{Antoniou:2019awm,Kleihaus:2019rbg,Kleihaus:2020qwo,Ibadov:2020btp}).
Their presence is due to the emergence of a node
at some value of the radial coordinate $\eta_\star$
in the determinant, that arises upon diagonalisation
of the ODEs.
These cusp singularities seem to be a rather common feature 
of scalarized wormholes.
Here we see, that they do not only arise for
GB theories, but are also present for CS theories.

The figures also show, that the domain of existence of
wormhole solutions increases strongly with 
increasing NUT charge. For the GB coupling,
the limit of vanishing NUT charge leads to the scalarized
wormhole solutions of Antoniou et al.~\cite{Antoniou:2019awm}.
For the CS coupling, the vanishing of the NUT charge needs special attention,
since in this case two branches of solutions arise,
as shown for the Schwarzschild-NUT solutions by Brihaye et al.~\cite{Brihaye:2018bgc}.
Along the first branch the coupling constant decreases with decreasing NUT charge,
analogous to the GB case.
However, along the second branch the coupling constant increases with decreasing NUT charge.
This is seen
in Fig.~\ref{fig_dom1}(b), where the solutions for $n=N/M=3$, 2 and 1 are on the first branch,
while the solutions for $n=N/M=0.5$ are already on the second branch.
Interestingly, the smaller the values of the NUT charge, the larger the values of the coupling constant
that are necessary to obtain scalarized solutions.  In fact, both conspire in such a way,
that a finite domain arises also in the CS case in the limit $N\to 0$
as further discussed below.

\subsection{Throat properties}

\begin{figure}[t!]
\begin{center}
\includegraphics[width=.45\textwidth, angle =0]{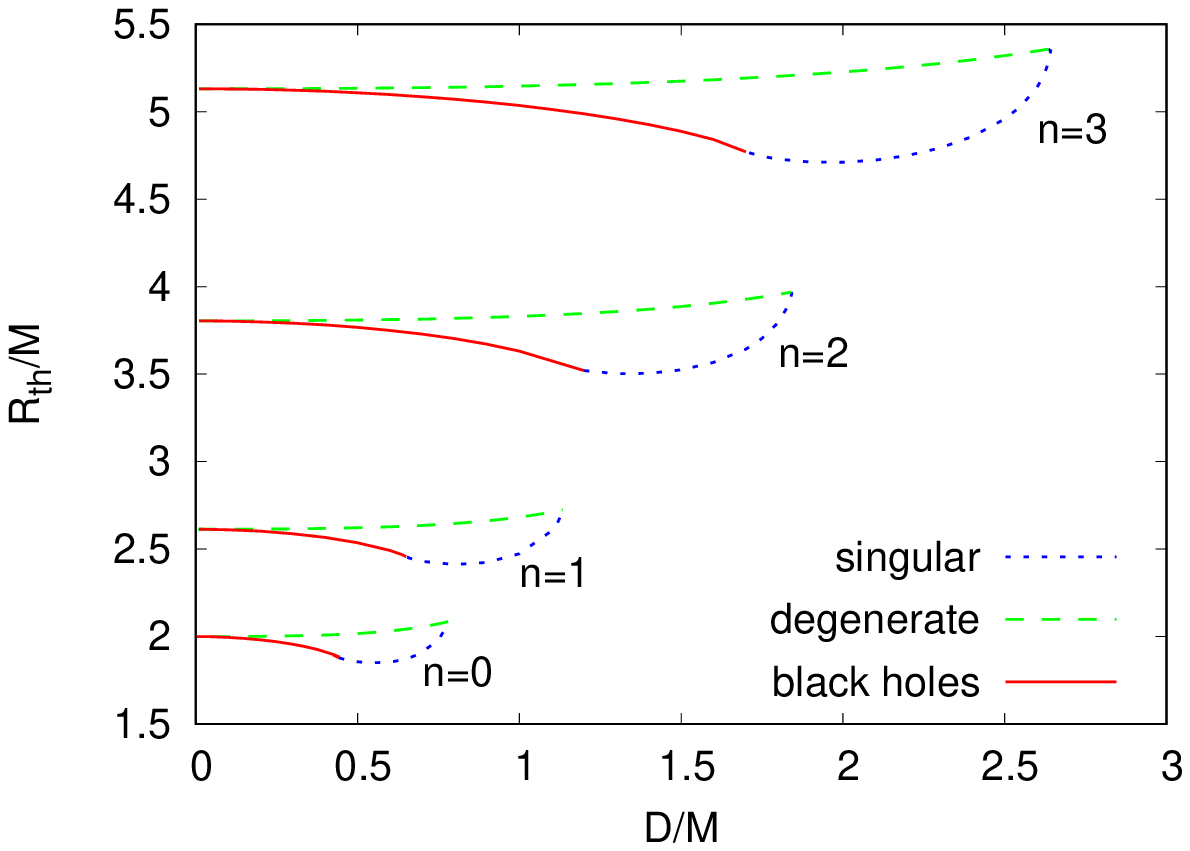}
\includegraphics[width=.45\textwidth, angle =0]{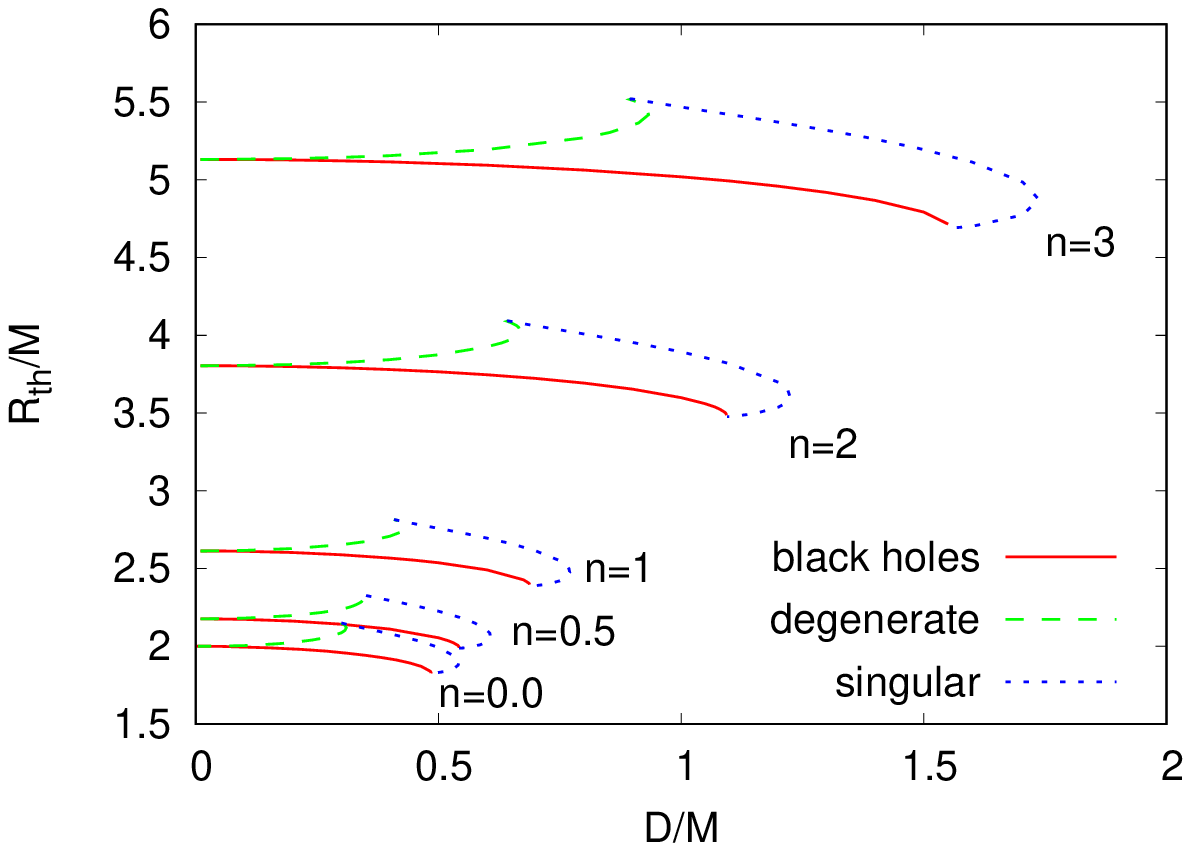}
\\
\includegraphics[width=.45\textwidth, angle =0]{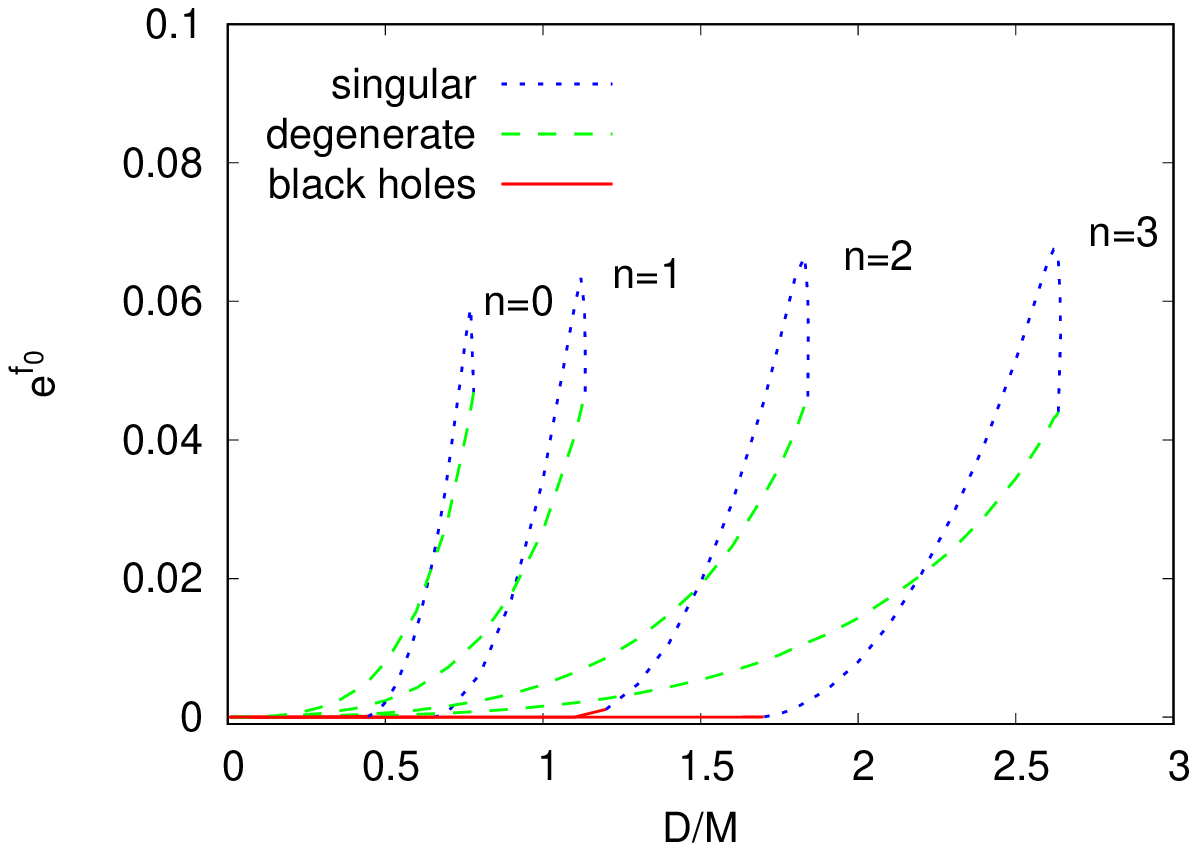}
\includegraphics[width=.45\textwidth, angle =0]{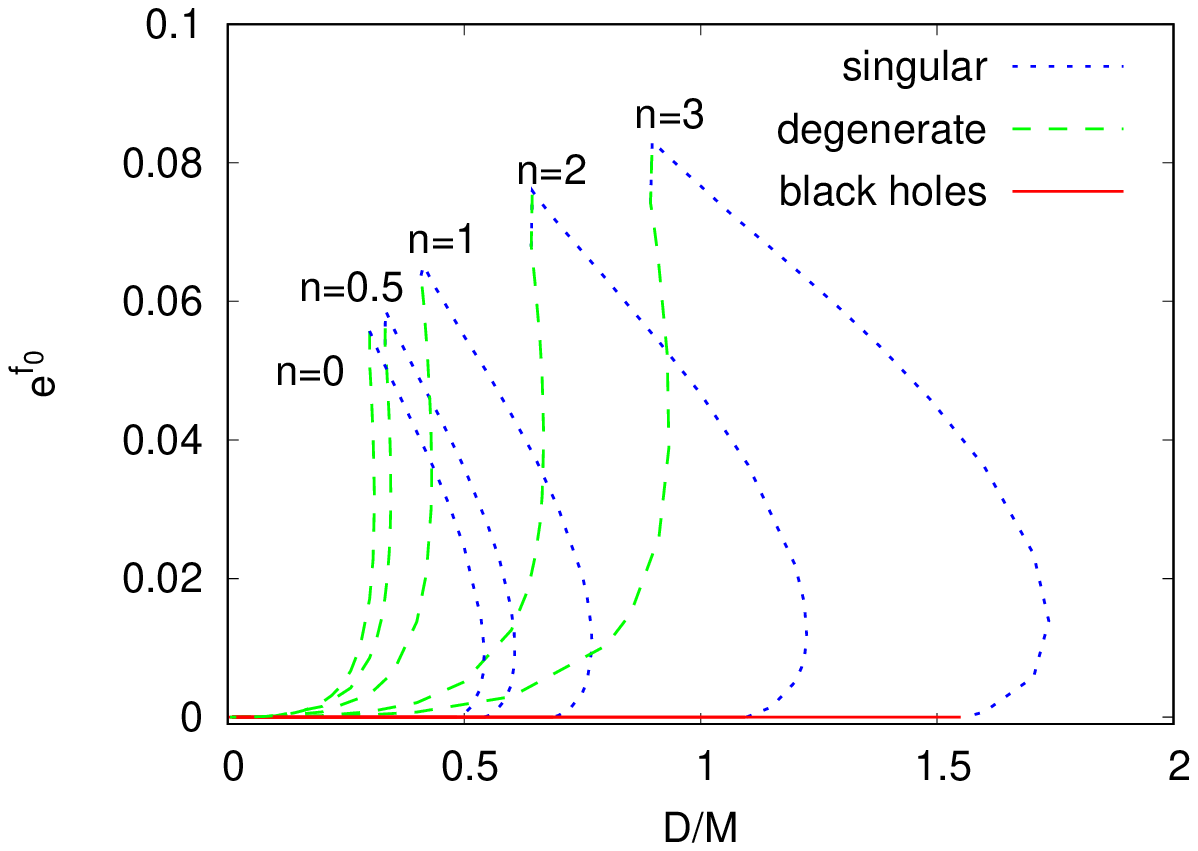}
\\
\includegraphics[width=.45\textwidth, angle =0]{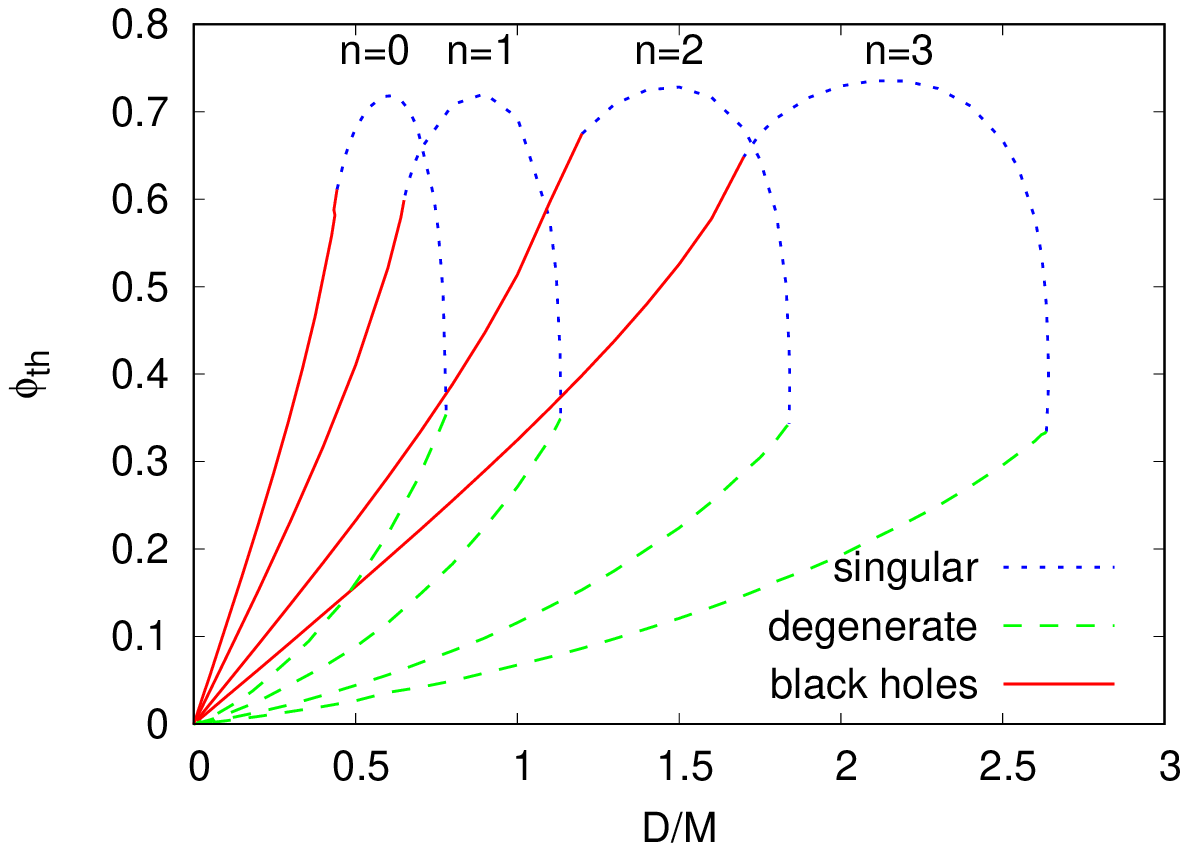}
\includegraphics[width=.45\textwidth, angle =0]{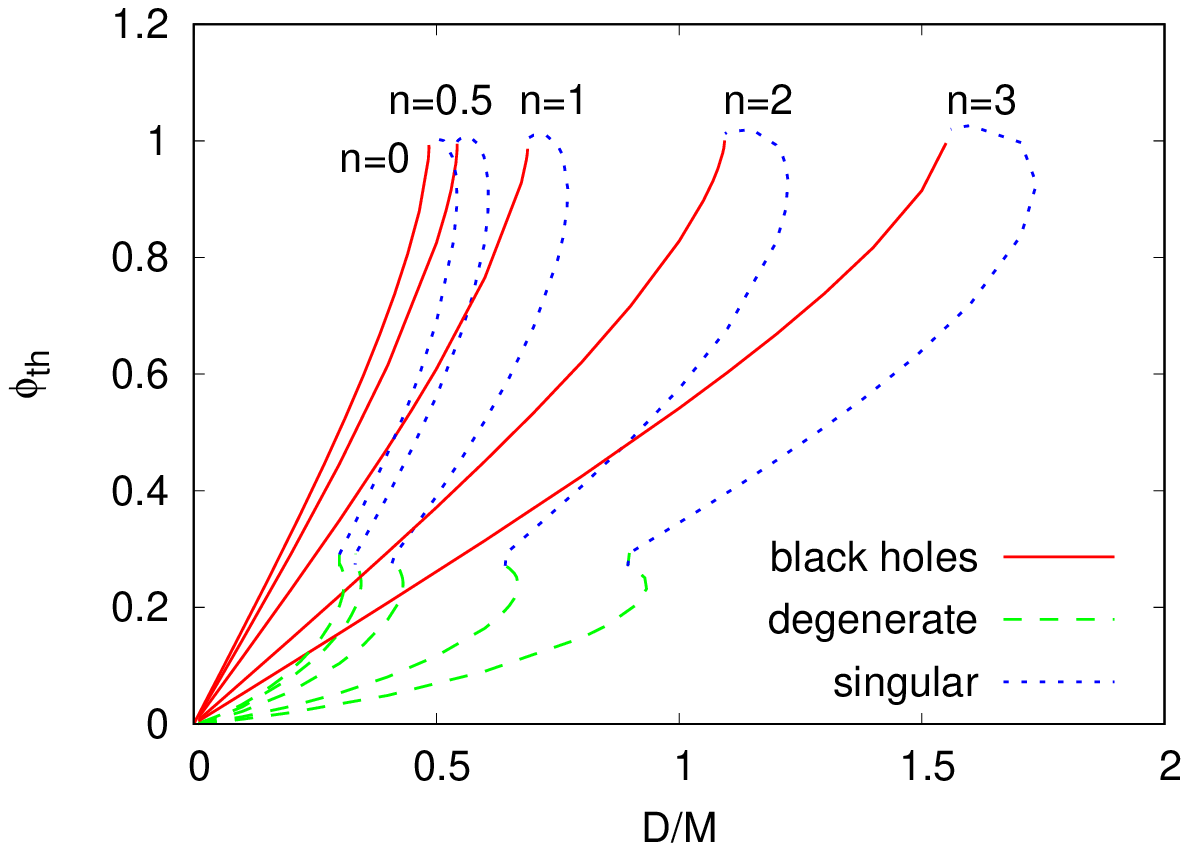}
\end{center}
\caption{
Properties at the throat
(left plots: Gauss-Bonnet, right plots: Chern-Simons)
for several values of the scaled NUT charge $n=N/M$:
(a) and (b) scaled circumferential radius $R_{\rm th}/M$,
(c) and (d) metric function $e^{f_0}$,
(e) and (f) scalar field $\phi_{\rm th}$ 
vs scaled scalar charge $D/M$.
The solid red curves represent the black hole limit,
the dashed green curves the degenerate wormhole limit,
and the dotted blue curves the singular limit.
}
\label{fig_dom2}
\end{figure}

We next address properties of the wormhole center,
and since we do not consider equators here, but only throats,
we refer to these properties as throat properties.
In particular, we consider the value of the circumferential radius $R_C$, Eq.~(\ref{rc}), at the throat
which we denote by $R_{\rm th}$, the value of the metric function $e^{f_0}$,
and the value of the critical angle $\theta_c$, Eq.~(\ref{thetac}).
For these properties we delimit their domains of existence in Fig.~\ref{fig_dom2},
by  the extracting the boundaries of the respective domains.
As before we show the black hole limit by solid red curves,
the degenerate limit by dashed green curves,
and the singular limit by dotted blue curves.

We exhibit the domain of the scaled value $R_{\rm th}/M$ versus the
scaled scalar charge $D/M$ in Fig.~\ref{fig_dom2}
for the GB invariant (a) and the CS invariant (b)
for several values of the NUT charge $n=N/M$.
We note that for a fixed value of the NUT charge $n=N/M$,
in the limit of vanishing scalar charge ($D/M\to 0$)
the same solution is
approached for both the GB invariant (a) and the CS invariant (b).
In this limit the scalar field vanishes identically, 
and thus there are no wormhole solutions.
The limiting solution therefore corresponds to
a Schwarzschild-NUT black hole solution, for which 
the value of $R_{\rm th}/M$ depends only on the value of the NUT charge $n=N/M$.

When taking the NUT charge to zero,
the boundary composed of scalarized black holes remains finite. 
In the GB case the scalarized Schwarzschild solutions are approached.
In the CS case the coupling constant diverges, as the NUT charge goes to zero,
leaving a finite value for the source term of the scalar field.
In fact a new coupling constant might be considered,
$\alpha'=\alpha/N$ such that $\alpha'$ is finite
and might be varied. 
Thus a peculiar set of limiting solutions arises,
that starts from the Schwarzschild black hole with $R/M=2$ at $D/M=0$
and forms the black hole boundary.

Considerung the full domain of existence, in the GB case in the limit $n=N/M \to 0$
the known finite domain of scalarized wormhole solutions is approached
\cite{Antoniou:2019awm}.
In the CS case the limit $n=N/M \to 0$ leads to a finite domain
of solutions as well, albeit solutions resulting from a  peculiar cancellation.
Nevertheless, the change of the domain of existence of the 
circumferential radius $R_{\rm th}/M$
is completely smooth in the limit $n=N/M \to 0$,
as calculations for several small values of $n=N/M$ have shown.
The limiting domain is seen in Fig.~\ref{fig_dom2}(b).
In this figure it does not make a difference whether the solutions
are on the first (large $n=N/M$) or second (small $n=N/M$) branch.

Inspection of the metric function $e^{f_0}$, shown
versus the scaled scalar charge $D/M$ in Fig.~\ref{fig_dom2}
for the GB invariant (c) and the CS invariant (d),
yields full agreement with the above discussion.
In the black hole limit the metric function vanishes as it must,
while the boundary of degenerate scalarized wormholes
approaches also the black hole limit
in the limit of vanishing scalar charge, $D/M \to 0$.
On the other hand, the boundary of singular wormholes connects to the
scalarized nutty black holes and contains the maximal value
of this metric function for a given value of the NUT charge $n=N/M$.
This maximum remains finite for $n=N/M \to 0$.

The scalar field at the throat $\phi_{\rm th}$ is exhibited
versus the scaled scalar charge $D/M$ in Fig.~\ref{fig_dom2}
for the GB invariant (e) and the CS invariant (f).
As argued above, for $D/M \to 0$ the scalar field vanishes,
and thus also $\phi_{\rm th} \to 0$.
Analogous to the metric function $e^{f_0}$,
for a given NUT charge $n=N/M$ a maximum of $\phi_{\rm th}$ is reached
on the singular boundary, and for $n=N/M \to 0$ the maximum
remains finite.

\subsection{Junction conditions and critical polar angle}

\begin{figure}[t!]
\begin{center}
\includegraphics[width=.45\textwidth, angle =0]{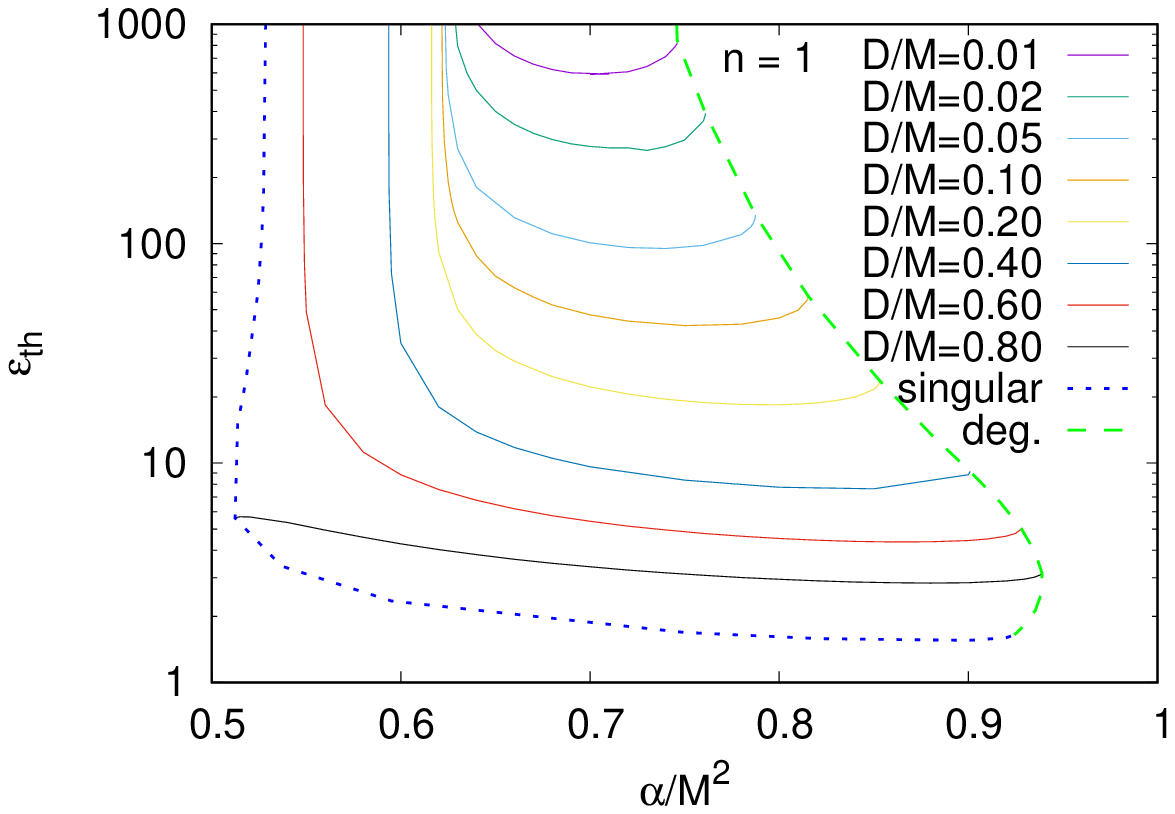}
\includegraphics[width=.45\textwidth, angle =0]{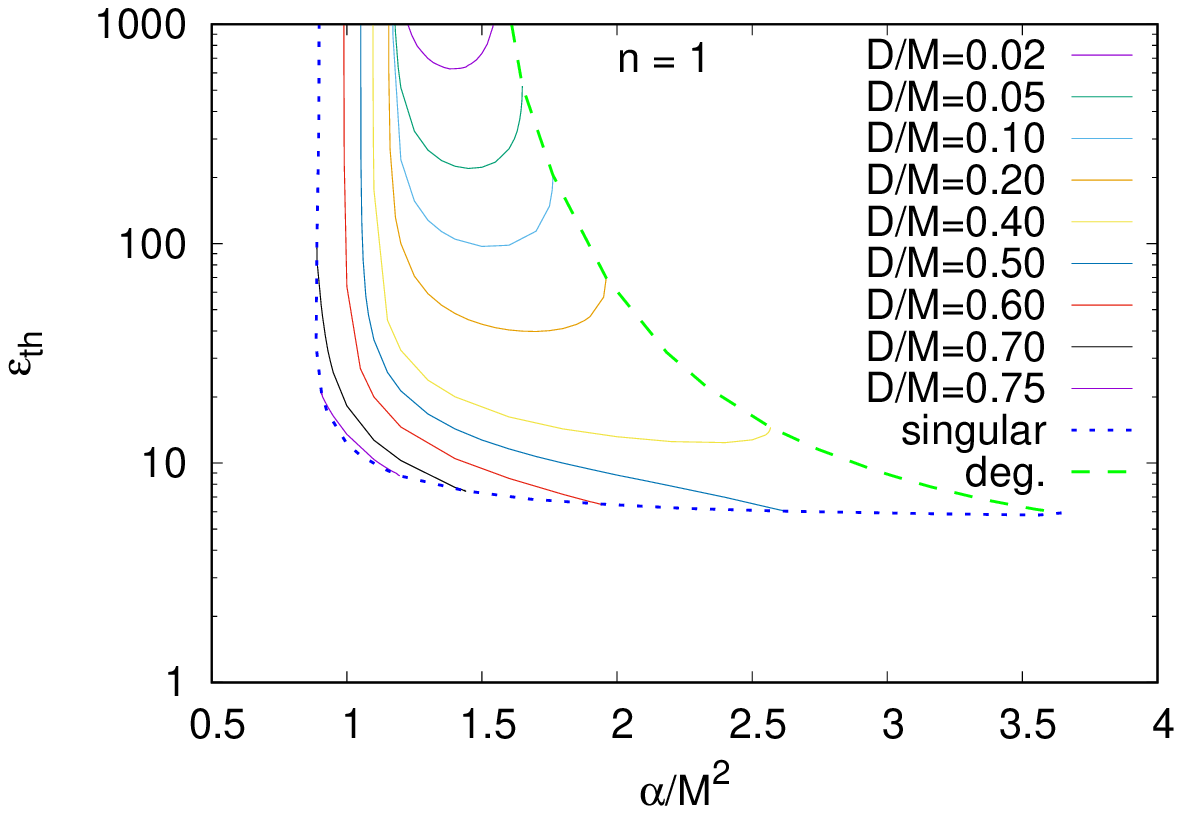}
\\
\includegraphics[width=.45\textwidth, angle =0]{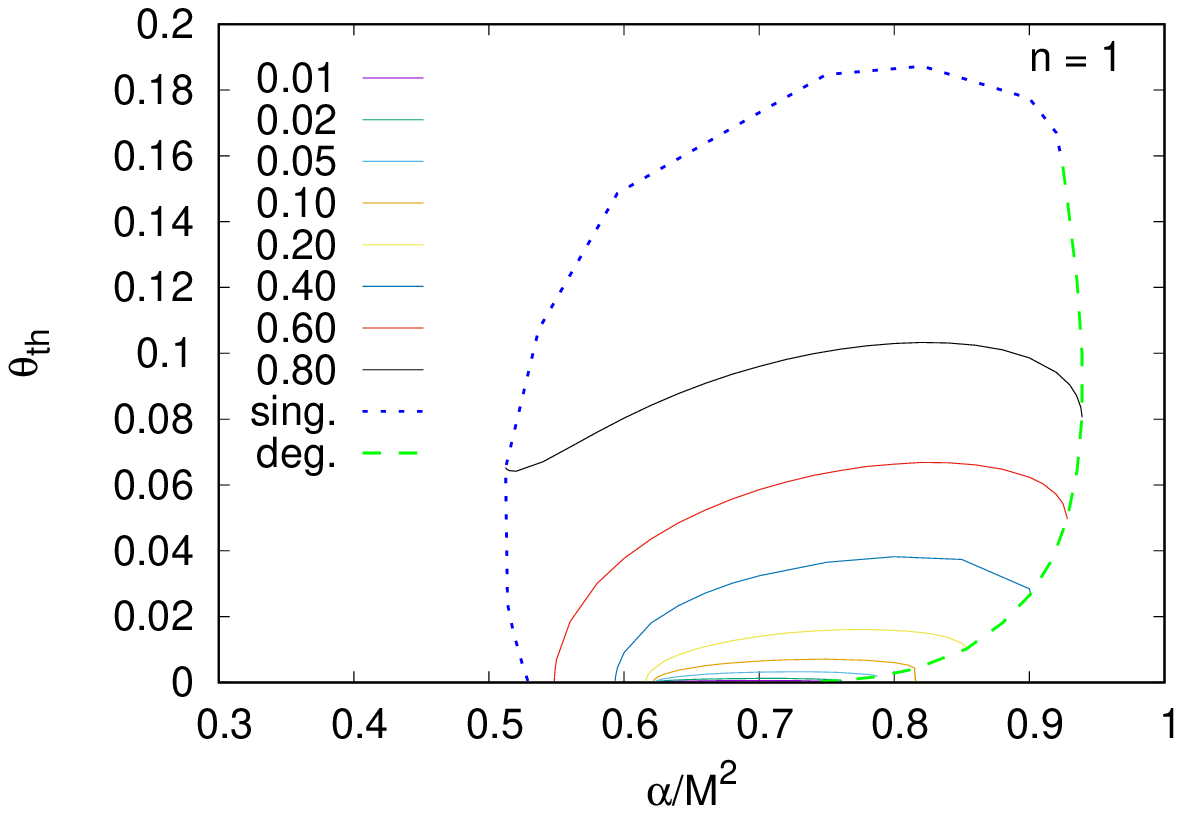}
\includegraphics[width=.45\textwidth, angle =0]{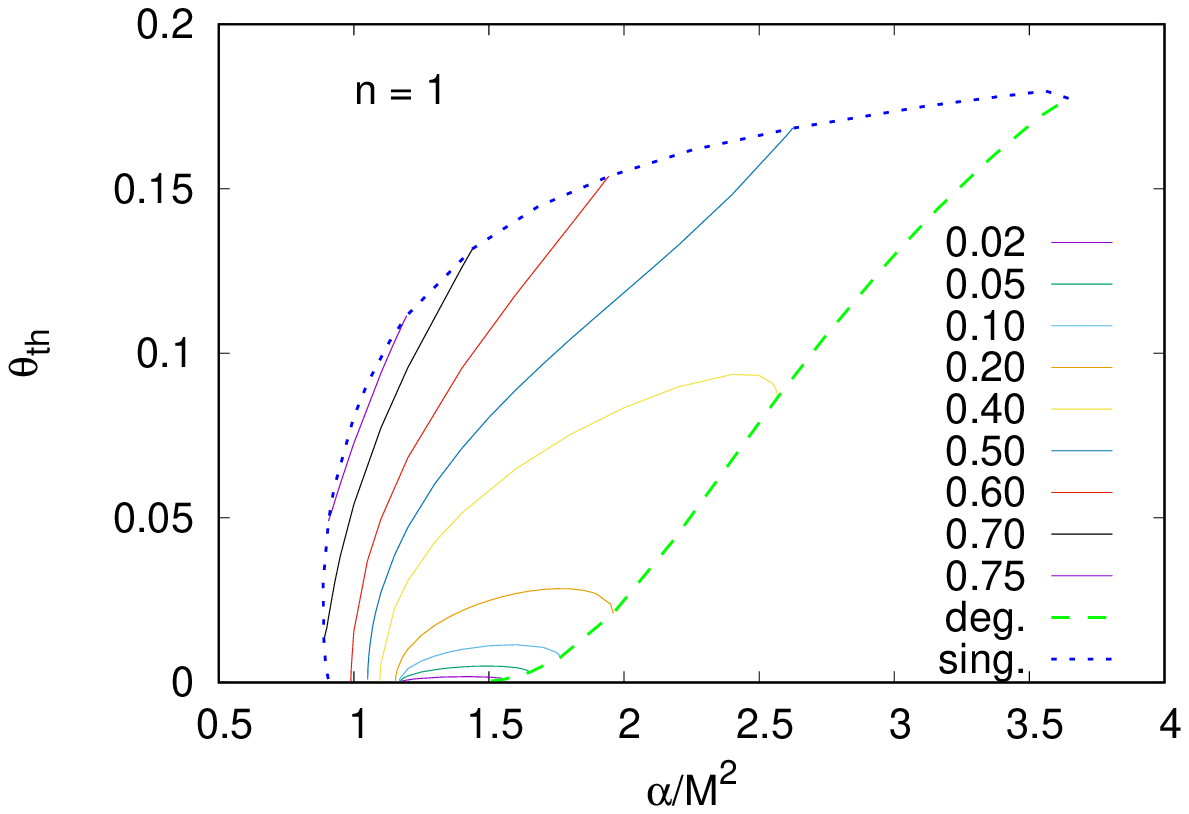} \end{center}
\caption{
Properties at the throat
(left plots: Gauss-Bonnet, right plots: Chern-Simons)
for scaled NUT charge $n=N/M=1$
and several values of the scaled scalar charge $D/M$:
(a) and (b) energy density $\epsilon_{\rm th}$,
(c) and (d) critical polar angle $\theta_{\rm th}$
vs scaled coupling constant $\alpha/M^2$.
The dashed green curves represent the degenerate wormhole limit,
and the dotted blue curves the singular limit.
}
\label{fig_dom3}
\end{figure}

As discussed above we need to satisfy junction conditions
in order to obtain symmetric wormholes without singularities
(except for the Misner string).
We have therefore introduced an action at the center
and allowed for a thin shell of matter in the form of a 
perfect fluid with energy density $\epsilon_c$.
Since we here focus only on wormholes with a single throat,
we denote this energy density by $\epsilon_{\rm th}$.
We exhibit in Fig.~\ref{fig_dom3} $\epsilon_{\rm th}$
for a pressureless fluid for several sets of wormhole solutions
with fixed values of the scaled scalar charge $D/M$
and NUT charge $n=N/M=1$ versus the scaled coupling constant $\alpha/M^2$
for the GB invariant (a) and the CS invariant (b).

The boundaries of these sets of solutions 
are given by the dashed green curves,
representing the degenerate wormhole limit,
and the dotted blue curves, representing the singular limit.
We note, that the energy density $\epsilon_{\rm th}$ is positive
in all cases shown. 
Thus these wormholes need only ordinary matter
at the throat to remain open.

Last we address the critical polar angle at the center $\theta_c$,
Eq.~(\ref{thetac}). 
Denoting the critical polar angle at the throat by $\theta_{\rm th}$,
we exhibit $\theta_{\rm th}$ also in Fig.~\ref{fig_dom3} 
versus the scaled coupling constant $\alpha/M^2$
for the same sets of solutions
for the GB invariant (a) and the CS invariant (b).
The critical angle reaches its maximal value of about
$\theta_{\rm th, max} \approx 0.19$ (a) and $ \approx 0.18$ (b)
on the boundary of singular wormhole solutions.
Along the boundary of degenerate wormholes 
the critical angle decreases monotonically,
until it vanishes, precisely when the Schwarzschild-NUT solution
is reached.

The horizon metric 
does not feature a finite critical polar angle for the Schwarzschild-NUT and the scalarized
nutty black holes.
For these black holes $\theta_{\rm th}=0$ 
only on the polar axis, $\theta=0$ and $\theta=\pi/2$.
Consequently, $\theta_{\rm th}=0$ along the black hole boundary.
Along the degenerate wormhole boundary $\theta_{\rm th}$ then increases
again. Thus all scalarized nutty wormholes do possess a finite
critical polar angle.
The region of closed timelike curves therefore extends
to the throat of these nutty wormholes.

\section{Conclusions}

Following the reasoning of Brihaye et al.~\cite{Brihaye:2018bgc},
who have studied spontaneouly scalarized Schwarzschild-NUT solutions,
whose scalarization is caused by the presence of either a GB term
or a CS term in the scalar field equation,
we have investigate scalarized nutty wormhole solutions
in these higher curvature theories.
The presence of a NUT charge leads to solutions with a Misner string
on the polar axis. However, the dependence of the polar angle
factorizes and thus only a set of coupled ODEs for the metric functions
and the scalar field arises. Moreover the usual boundary conditions
at spatial infinity are retained. 

Solving numerically the ODEs we have obtained scalarized nutty wormhole solutions
for both higher curvature invariants.
These wormhole solutions possess a minimum of the circumferential radius,
that arises naturally in these solutions, and that is identified with the wormhole throat.
When integrating the ODEs beyond the throat, a maximum may be encountered,
that would correspond to an equator. In any case, when integrating further
into the second part of the manifold a singularity would be encountered at some
value of the radial coordinate.
To avoid such a singularity, we have imposed reflection symmetry on the
wormhole solutions at the throat, and satisfied the resulting junction conditions
in terms of an action with a thin shell of ordinary matter at the throat.

The novel scalarized GB wormholes represent a NUT generalization of the
previously obtained scalarized wormholes 
\cite{Kanti:2011jz,Kanti:2011yv,Antoniou:2019awm,Ibadov:2020btp}.
In contrast, scalarized CS wormholes were not studied before, 
since a finite CS term is necessary for their existence,
while a spherically symmetric metric leads to a vanishing CS invariant.
Nevertheless the scalarized nutty wormholes with both invariants show many features
already known from the ordinary scalarized GB wormholes.
In particular, their domains of existence contain the same type of boundaries.
These boundaries consist of the respective black hole boundary,
of a boundary, where singular solutions are reached,
and of a degenerate boundary,
where two extrema, the equator and the throat, merge.
However, the NUT charge gives rise to regions of the spacetime, 
where closed timelike curves exist, that extend to the throat.

Considering the limit of vanishing coupling constant,
the Schwarzschild-NUT solutions 
are obtained for both invariants. Then wormhole solutions do not exist
any longer, since they need the higher order curvature terms in
the (generalized) Einstein equations to obtain an effective stress energy tensor
that violates the NEC conditions.
Considering on the other hand the limit of vanishing NUT charge,
for both invariants a finite domain of existence of scalarized solutions results.
For the GB term this is no surprise, since their existence was shown before
\cite{Kanti:2011jz,Kanti:2011yv,Antoniou:2019awm,Ibadov:2020btp}.
For the CS, however, this limit is special.
Scalarization needs higher and higher values of the coupling constant,
as the NUT charge is lowered towards zero.
In the limit, the coupling constant diverges as the NUT charge goes to zero.
In that case, the CS term approaches a finite limiting value,
which gives rise to a finite limiting domain of existence.
Note that the spacetime of these wormhole and black hole solutions do not
possess a Misner string anymore and consequently no closed timelike curves exist.

The presence of a NUT charge can be viewed as toy model
for learning about scalarized rotating wormhole solutions.
Their construction still represents a challenging task,
not only because complicated sets of coupled partial differential equations
need to be solved, but also because the proper conditions for the throat
need to be formulated together with the proper set of junction conditions.
However, also further directions of research look promising
like, in particular, the inclusion of further fields
(see e.g., \cite{Clement:2015aka,Brihaye:2020dgo}).


\section*{Acknowledgement}

BK and JK gratefully acknowledge support by the
DFG Research Training Group 1620 {\sl Models of Gravity}
and the COST Action CA16104. 


\end{document}